\documentclass{article}
\usepackage[utf8]{inputenc}
\usepackage[letterpaper]{geometry}
\usepackage{libertine}
\usepackage{libertinust1math}
\usepackage[T1]{fontenc}
\usepackage{amsmath}
\usepackage{amsfonts}
\usepackage{amsthm}
\usepackage{soul}
\usepackage{url}
\usepackage[]{authblk} 
\usepackage{datetime}
\usepackage{hyperref} 
\usepackage{graphicx}
\usepackage{array}
\usepackage{multirow}
\usepackage{float}
\usepackage{csquotes}
\usepackage{subfig}

\newcommand{\PreserveBackslash}[1]{\let\temp=\\#1\let\\=\temp}
\newcolumntype{C}[1]{>{\PreserveBackslash\centering}p{#1}}

\newdateformat{monthyear}{\monthname[\THEMONTH] \THEYEAR}
\newcommand*{\email}[1]{
    \normalsize\href{mailto:#1}{#1}
    }
\title{Artificial Intelligence for Social Good: A Survey}
\author{Zheyuan Ryan Shi, Claire Wang, Fei Fang}
\affil{\normalsize Carnegie Mellon University\\ \email{ryanshi@cmu.edu}, \email{clairew1@andrew.cmu.edu}, \email{feif@cs.cmu.edu}}
\date{\normalsize First draft: \monthyear\today}

\begin{document}

\maketitle

\begin{abstract}
    Artificial intelligence for social good (AI4SG) is a research theme that aims to use and advance artificial intelligence to address societal issues and improve the well-being of the world. AI4SG has received lots of attention from the research community in the past decade with several successful applications.
    Building on the most comprehensive collection of the AI4SG literature to date with over 1000 contributed papers, we provide a detailed account and analysis of the work under the theme in the following ways. (1) We quantitatively analyze the distribution and trend of the AI4SG literature in terms of application domains and AI techniques used. (2) We propose three conceptual methods to systematically group the existing literature and analyze the eight AI4SG application domains in a unified framework. (3) We distill five research topics that represent the common challenges in AI4SG across various application domains. (4) We discuss five issues that, we hope, can shed light on the future development of the AI4SG research.
\end{abstract}
\section{Introduction}
Artificial intelligence (AI) is in many ways an important source of change in people's lives in the 21st century. Its impact is drastic and real: Youtube's AI-driven recommendation system would present sports videos for days if one happens to watch a live baseball game on the platform~\cite{RecSys:2016:Covington:youtube}; email writing becomes much faster with machine learning (ML) based auto-completion~\cite{KDD:2019:Chen:gmail}; many businesses have adopted natural language processing based chatbots as part of their customer services~\cite{ARXIV:2015:Vinyals:conversation}. AI has also greatly advanced human capabilities in complex decision-making processes ranging from determining how to allocate security resources to protect airports~\cite{AAMAS:2008:Pita:laairport} to  games such as poker~\cite{Science:2018:Brown:poker} and Go~\cite{Science:2018:Silver:alphazero}. All such tangible and stunning progress suggests that an ``AI summer'' is happening. As some put it, ``AI is the new electricity''~\cite{NEWS:2017:Lynch:aielectricity}.

Meanwhile, in the past decade, an emerging theme in the AI research community is the so-called ``AI for social good'' (AI4SG): researchers aim at developing AI methods and tools to address problems at the societal level and improve the well-being of the society. Over the years, there have been several successful AI4SG projects, such as guiding municipal water pipe replacement~\cite{KDD:2018:Abernethy:flint}, protecting wildlife from poaching~\cite{IAAI:2016:Fang:gsg}, and spreading HIV prevention information among homeless youth~\cite{AAMAS:2017:Yadav:hiv}. AI4SG has received recognition from both within and outside the academic community. Major AI conferences have featured various special tracks and workshops dedicated to AI4SG, with over 1000 papers on AI4SG topics formally published. Large companies are expanding their investments on AI4SG initiatives. The need for collaboration among researchers, governments, companies, and non-profit organizations to solve AI4SG problems has never been more widely appreciated.

\paragraph{What exactly is AI for social good?}
As we celebrate the progress of AI and AI4SG, let us take a step back to ponder what AI4SG really means. AI4SG is a vague concept, given that neither AI nor social good has a widely accepted definition.
Some of the recent efforts try to define AI4SG based on the realized or potential social impact and non-profit nature.
However, there is no consensus on such a definition.
One can argue that ``social impact'' is both too exclusive and inclusive: lots of AI4SG research has not (yet) achieved any tangible social impact, and AI research that does have some social impact may not be \textit{doing good}. The ``non-profit nature'' is similarly debatable since for-profit industry companies arguably contribute a lot to AI4SG, and the majority of research on AI and transportation, which most people would count in AI4SG, is not always intended for non-profit applications. Currently, the most popular set of efforts is to describe  AI4SG by listing the application domains of AI technologies, or even simpler, referring to ``societal challenges, which have not yet received significant attention by the AI community''.\footnote{\url{https://aaai.org/Symposia/Spring/sss17symposia.php}} As Berendt points out, social good (common good) is referred to as a goal, but not defined~\cite{DG:2019:Berendt:ai4sg}.

We do not attempt to propose any new definition of AI4SG. In this survey, we adopt the widely used approach by enumerating several application domains. This is not perfect, as we will unavoidably miss some works.
However, by doing this we are able to provide a systematic overview of the field, while the community continues to discuss and iterate on a better definition.
In fact, we believe that the lack of precise definition might be more of a strength for the development of AI4SG than a problem. Just like the lack of common definition for AI helped the field to grow and innovate beyond its boundary, we think an inclusive boundary of AI4SG could encourage more researchers to contribute to this area.

\paragraph{Why write this survey}
AI4SG as a research theme has grown tremendously in the past decade. We believe this is the time to summarize the progress and identify the research and non-technical challenges so far.
The existing surveys or survey-like papers are inadequate in the following four aspects which we attempt to address.

First, the literature lacks a comprehensive survey of this domain.
Berendt provides a survey of 99 AI4SG papers, focusing on publications from a few (mostly data science) workshops~\cite{DG:2019:Berendt:ai4sg}.
The CCC report~\cite{CCC:2017:Hager:ai4sg} and Cowls et al.~\cite{NA:2019:Cowls:ai4sg} cover a few successful projects from the highly regarded venues, yet lack the breadth of coverage. Chui et al.~\cite{McKinsey:2018:Chui:ai4sg} built a collection of 160 AI4SG projects, yet these projects are not explicitly discussed in the paper. We cover over 1000 AI4SG publications at major AI venues in the past 12 years and provide a detailed account of the research in eight major application domains.

Second, we provide the first quantitative analysis of the AI4SG research projects. We show the trend of each application domain and AI technique in terms of the number of papers published. We also identify the hot spots that have attracted the most research effort in the current AI4SG community.

Third, most of the current literature discusses the AI4SG literature according to application domains without a systematic approach to analyze and categorize AI4SG research. We use three conceptual methods to  group the existing works and to dissect each application domain. These categorizations yield insights into how one can identify concrete ``social good'' problems that AI can be used to address.

Fourth, existing works do not adequately draw the two-way interaction between the technical advances in AI research and the applications. In addition to using advances in AI for societal applications, we emphasize that applications in many different domains also motivated advances in some common meaningful research directions. We distill several technical research topics stemmed from the nature of AI4SG research.

Nonetheless, we leave out a few aspects of AI4SG which are already covered in previous works. The interested reader might refer to the work of Berendt~\cite{DG:2019:Berendt:ai4sg} for a more detailed discussion on ``what is social good, are we really doing good, and how to ensure that we do good?''. Chui et al.~\cite{McKinsey:2018:Chui:ai4sg} have more discussion on the action plan for those outside the academic research community.

\paragraph{Target audience}
This survey is primarily targeted at researchers in the AI research community who desire to make a positive and tangible impact in the social good domains.

AI4SG has a wide range of stakeholders: AI researchers who aspire to make positive social impact, researchers in other domains who hope to leverage the promise of AI, practitioners who provide the ``infrastructure'' in place and who make everything in this community possible, funding agencies which decide on grant and resources allocations for AI4SG projects, companies which want to utilize their talent and resources in-house to promote corporate social responsibility, media that shape the public discourse of AI and its positive (and negative) impact on our society, and most importantly, everyone who will be affected by the AI4SG projects, and that is, everyone.

Our survey is mainly tailored to the AI research community. However, we believe that the present paper will also be informative to other stakeholders of AI4SG. In particular, we show in the paper the current development of AI4SG, the application domains and specific problems that have been studied, and the AI4SG projects that have been successfully deployed.

\paragraph{The scope of this survey}
In this survey, we focus on the AI technologies which are developed within the AI communities with the primary purpose of producing \textit{positive} outcomes for inherent societal problems.

As AI technologies become increasingly powerful and democratized, we note several unfortunate cases where AI is developed for outright malicious use, e.g. DeepNude~\cite{website:2019:deepnude}, or for highly dubious purposes~\cite{EMNLP:2019:Yang:newscomment}. We believe such cases represent an urgent and complex problem that the AI community, and much more beyond the AI community, need to address together. 
Such malicious use of AI is equally important as using AI to produce positive outcomes. In this survey, we focus on the latter and refer the reader to the work of Brundage et al.~\cite{NA:2018:Brundage:maliciousai} for detailed coverage of the malicious use of AI.

We leave out the emerging literature on the fairness, accountability, transparency, and ethics problems of AI solutions to social good problems, which is surveyed in a few recent papers~\cite{NA:2019:Mehrabi:fairness,DSAA:2018:Gilpin:interpretabilityml,Oxford:2019:Kearns:ethical}. In the present work, we focus on inherent societal problems (including, e.g. inequality and fairness) but not the ``second-order problems'' about the AI solutions. However, we do believe these issues are critical to the successful delivery of AI4SG solutions and we encourage the community to make more progress.

AI and public policy is another important area related to AI4SG. The reader may refer to the AI Now report~\cite{NA:2019:Crawford:ainow} for a detailed treatment of the current topics.

The majority of works surveyed in this paper originate from (or partially from) the AI community. We recognize that the use of AI techniques has a growing presence in disciplines other than computer science. For each application domain, we will briefly discuss the relevant literature from within the domain, where applicable.

\paragraph{Outline}
This survey is organized as follows. In Section~\ref{sec:overview}, we provide an overview of the AI4SG literature. We characterize the distribution and trend of AI4SG research over application domains and AI techniques. In Section~\ref{sec:application}, we review the eight main application domains of AI4SG research. For each domain, we introduce the key topics and challenges to which AI has been applied, and a case study on a successfully deployed AI4SG project. In addition to applying AI techniques to societal problems, it is also part of AI4SG research that the experience from application motivates and leads to advances in several key AI research directions. In Section~\ref{sec:research}, we distill five such AI research topics which have their root across multiple application domains. Section~\ref{sec:discussion} features a discussion of several important aspects of AI4SG research such as evaluation and deployment. We summarize the paper in Section~\ref{sec:conclusion}.

\section{Overview of the AI for Social Good Literature}
\label{sec:overview}

\subsection{Literature Selection}
For this survey, we selected 1176 papers from the AAAI,\footnote{The AAAI Conference on Artificial Intelligence} IJCAI,\footnote{International Joint Conference on Artificial Intelligence} IAAI,\footnote{Conference on Innovative Applications of Artificial Intelligence} AAMAS,\footnote{International Conference on Autonomous Agents and Multiagent Systems} KDD,\footnote{ACM SIGKDD International Conference on Knowledge Discovery \& Data Mining}  and COMPASS\footnote{ACM SIGCAS Conference on Computing and Sustainable Societies} conferences from the years 2008-2019, to analyze the trend of AI4SG research. Starting from Section~\ref{sec:application}, we discuss over 400 papers in detail, introducing the problems they address and the AI techniques being used. AAAI and IJCAI are general conferences on AI, which capture most of the relevant sub-topics in AI. AAMAS and KDD feature works in multi-agent systems and data mining, two sub-topics of AI that have lots of AI4SG related works. Each of these four conferences has some application-related track or session, which is exactly the type of paper we wish to focus on. IAAI is a dedicated venue for the deployed or emerging applications of AI in the real world. We chose COMPASS for its focus on ``computing and sustainable societies''. Although it is not an exclusively AI-focused venue, it nevertheless has an ample amount of AI- and AI4SG-related work.
In a future version, we also plan to include AI4SG-related works at conferences on other sub-topics of AI, such as natural language processing (e.g. ACL\footnote{Meeting of the Association for Computational Linguistics}) and computer vision (e.g. CVPR\footnote{IEEE/CVF Conference on Computer Vision and Pattern Recognition}).

For each of AAAI, IJCAI, and AAMAS, we went through their proceedings and programs on their respective websites for the years 2008-2019 and looked for any relevant sections to the topic of AI4SG. Specifically, we looked for sections with titles matching any of the following: 
Multidisciplinary Applications, Applications, Machine Learning Applications, Computational Sustainability, AI for Social Impact/Social Good, AI and Humans, or AI for Improving Human Well-being. For other venues, including KDD, IAAI, and COMPASS, we took all papers from the proceedings. 

After collecting all the papers, we then came up with initial sets of 20 keywords or phrases for each of the eight following domains: agriculture, education, environmental sustainability, healthcare, combating information manipulation, social care and urban planning, public safety, and transportation. 
In particular, information manipulation includes challenges such as misinformation and reputation fraud; social care and urban planning include topics such as public services and crisis management.
We also created similar keyword sets for each of the 16 following AI sub-topics: computer vision (CV), robotics, planning, routing and scheduling, multiagent systems, natural language processing (NLP), knowledge representation and reasoning, human computation and crowdsourcing, constraint satisfaction and optimization, game playing and interactive entertainment, game theory and economic paradigms, heuristic search and optimization, cognitive modeling, cognitive systems, reasoning under uncertainty, human-AI collaboration, and machine learning (ML). We chose these 16 sub-topics according to the keywords of AAAI-19.\footnote{\url{https://aaai.org/Conferences/AAAI-19/aaai19keywords/}} We then tagged each paper with any domain areas and techniques for which the title or abstract contained a keyword corresponding to those domains/techniques, making use of $\texttt{nltk.stem}$. 

After performing this keyword matching once for all papers collected, we manually looked through the abstracts of all papers that were not tagged with any domain area and, if we deemed them relevant to one of our AI4SG domains, added some salient keywords from the abstract to our collection of keywords. After this manual review, we re-ran the keyword matching using the updated keyword base. (We also briefly did the same manual inspection with papers without a technique tagging, though for such papers it became clear that they simply did not reference any technical attributes in the abstract or title.) 
After the updated keyword matching, we manually inspected the matched papers and remove those that are not actually relevant to the domain.
The collection of papers and the script can be found in our GitHub repository.\footnote{\url{https://github.com/AIandSocialGoodLab/AI4SG-Survey}}

\subsection{Distribution and Trend}
\begin{figure}
\centering
\subfloat[Evolution of AI4SG application domains]
{\includegraphics[width=0.9\columnwidth]{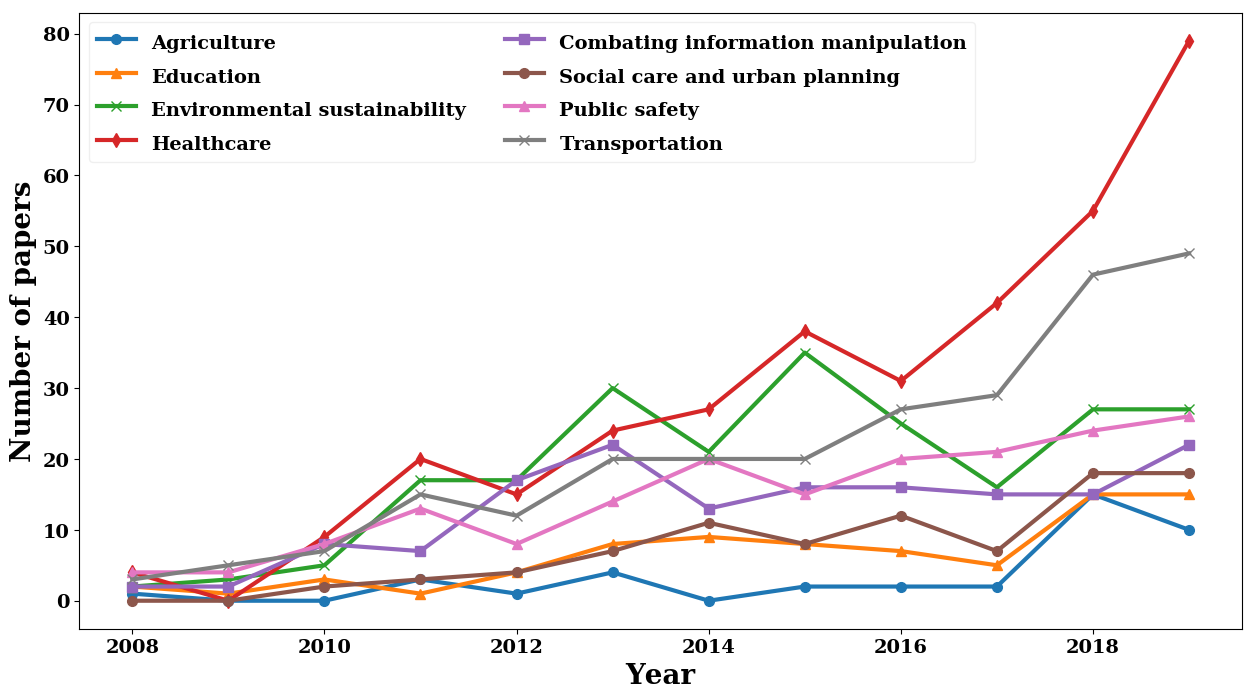} \label{fig:domaintrend}}\\
\subfloat[Evolution of AI4SG AI techniques]
{\includegraphics[width=0.9\columnwidth]{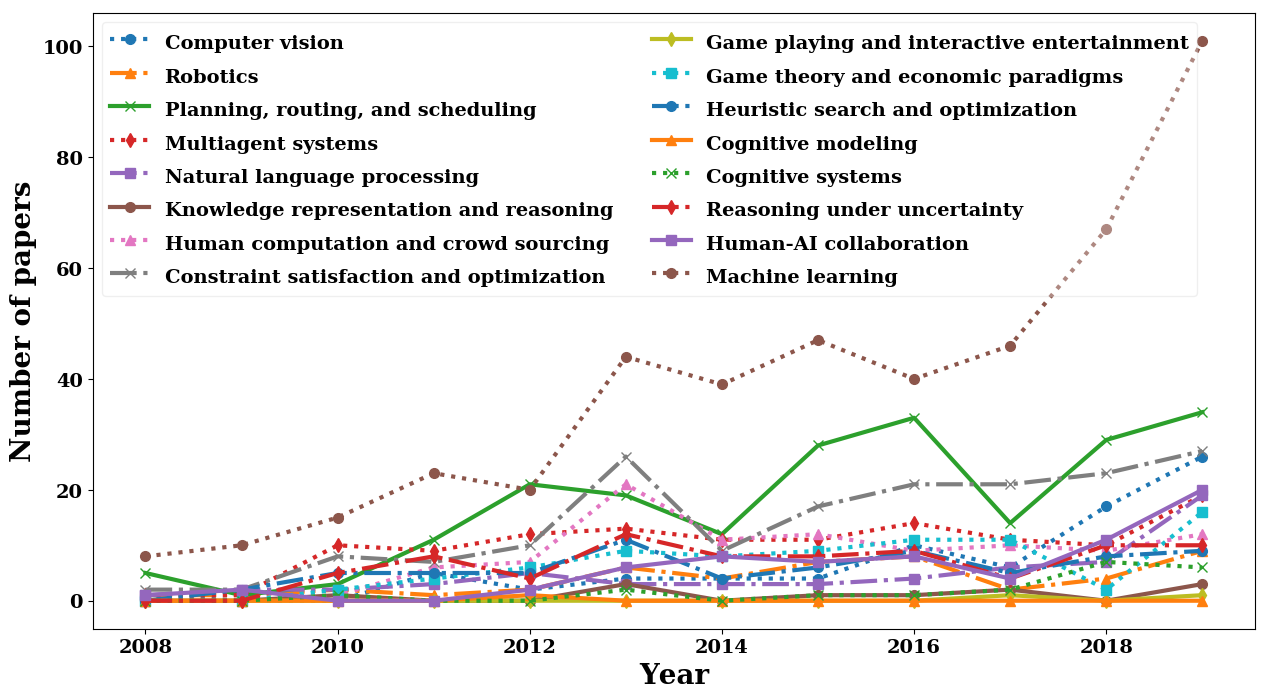} \label{fig:techniquetrend}}
\caption{Trend of AI4SG publications in different domains and with different techniques.}
\label{fig:trend}
\end{figure}

First, we summarize our findings in Fig.~\ref{fig:trend}. In general, across all domains and techniques, the number of papers pertaining to any of our listed AI4SG domains increased over time. In 2008, there were 18 papers on AI4SG and in 2019 we have 246, suggesting that research interest in AI4SG has been generally increasing in recent years.

As shown in Fig.~\ref{fig:domaintrend}, all domains have generally witnessed an upwards-tending trend in the number of papers published. Healthcare has received the most attention among all application domains. The difference between it and other domains appears to be widening, with the works on healthcare count 32\% of the total literature on AI4SG in 2019.
Transportation has a steady increase in publications and ranks second in the total number of papers in the past years. On the other hand, agriculture, education, and social care and urban planning have been the topics with the smallest literature across the years. This trend is likely due to the fact that both transportation and healthcare are well-established disciplines even outside of the computational realm. Many research topics within these areas are already mature without the presence of AI techniques. It is also relatively easier to gather data in these domains, with existing architecture such as electronic health records and traffic monitoring systems. Furthermore, the data pertaining to transportation and healthcare are usually in a precise, structured format that can be fed directly to an AI algorithm. On the other end of the spectrum, data pertaining to realms like social care and education are often in forms like interviews and questionnaires. Such data are harder to both collect and parse. Further, these two domains, along with that of agriculture, do not have well-established existing frameworks for data collection to begin with, further exacerbating the lack of data for analysis and usage with AI algorithms.

Up until 2016, environmental sustainability was among the most popular domains in AI4SG research, until transportation surpassed environmental sustainability. This could potentially be explained by a tendency for research done in the transportation domain being ultimately motivated by sustainability and energy efficiency.  Interestingly, around 2013 the domains of healthcare, combating information manipulation, and transportation all had roughly the same frequencies; after that, healthcare saw a steep increase in frequency, perhaps explained by recent advancements in the industry.

In Fig.~\ref{fig:techniquetrend} we show the trend of AI techniques in the AI4SG literature.
ML has been a consistently dominant technique in AI4SG papers, and its usage has surpassed that of other techniques even more in recent years (since 2013). This comes with little surprise, especially as keywords relating to ML techniques are quite generic and may encompass other, more specific techniques. For example, most of the recent work in computer vision (CV) and natural language processing (NLP) is built upon ML techniques. Planning, routing, and scheduling and constraint satisfaction and optimization have both been on a slower and somewhat unsteady upwards trajectory in the number of papers through the years. On the other hand, techniques like cognitive modeling, game-playing and interactive entertainment, and knowledge representation have stayed nearly stagnant in terms of the rate of publication over the years.

\begin{figure}[t]
  \centering
    \includegraphics[width=\textwidth]{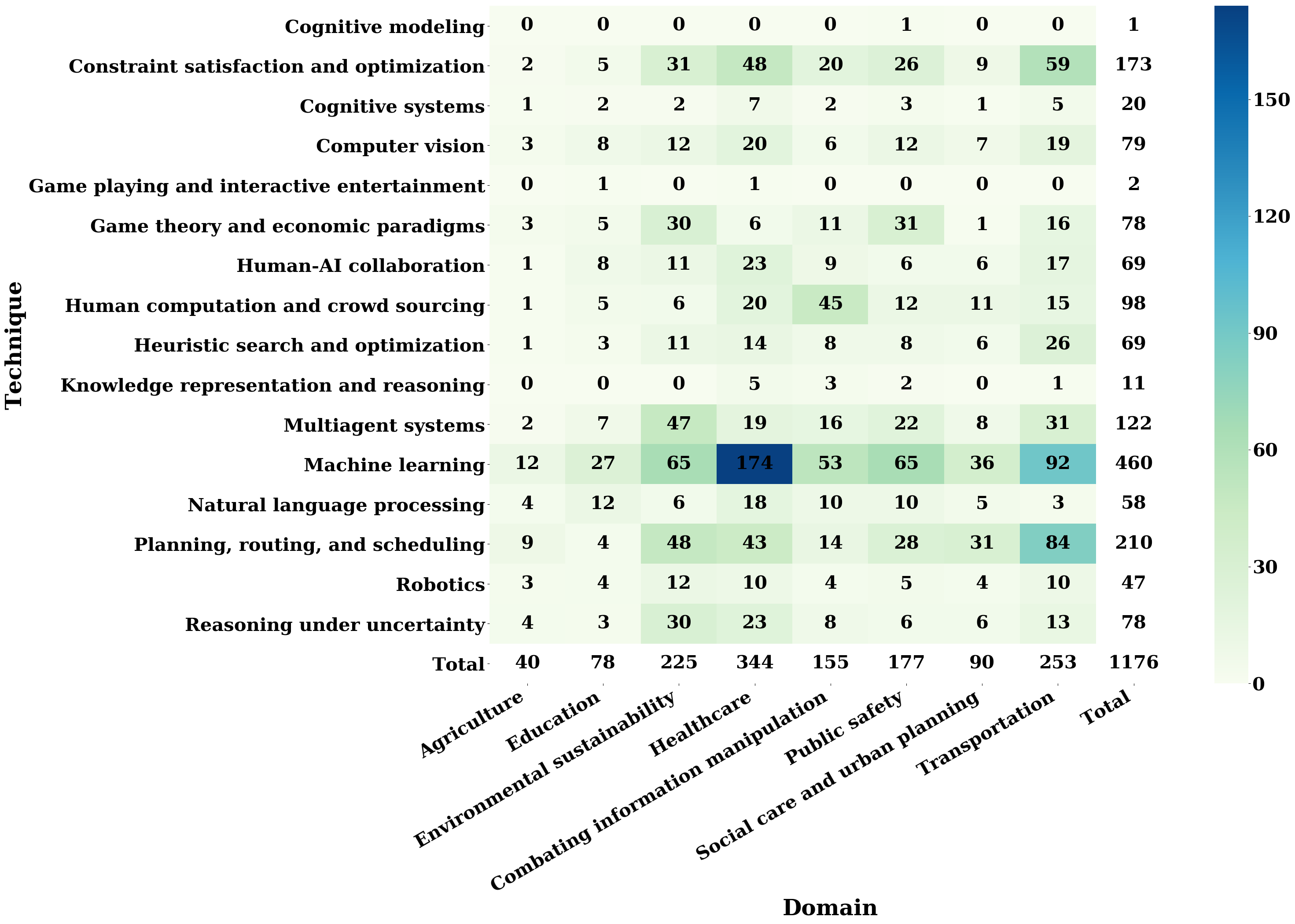}
  \caption{Heatmap of AI4SG paper domains and techniques.}
  \label{fig:heatmap}
\end{figure}

Fig.~\ref{fig:heatmap} shows a heatmap that visualizes the connection between application domains and AI techniques, in which we plot the number of papers falling into each (technique, domain) pair. We note that the numbers for each application or technique are not equal to the summation of numbers in the respective column/row since each paper might belong to multiple application domains and techniques.

The heatmap again reveals the prevalence of ML techniques, regardless of the domain. Techniques like game-playing and interactive entertainment, cognitive modeling, and knowledge representation are very infrequently utilized.
We also see again that application domains like transportation, healthcare, public safety, and environmental sustainability are the most frequently studied in AI4SG papers. 

The hottest combination is unsurprisingly machine learning in healthcare with 174 papers which surpasses all other combinations by a large margin.
Within the transportation domain, planning and constraint satisfaction are frequently utilized techniques. This makes sense given transportation's natural domain properties of being constrained by a number of dynamic variables and needing to plan sequences of actions. Within public safety, game theory is a popular approach since it characterizes the adversarial nature of many problems in this domain.
Another hot combination that stands out is the usage of human computation techniques towards combating information manipulation. This is another natural pairing, may be explained by the ease of analyzing information in mass media via the many consumers of such information. 

\section{Research Progress in AI for Social Good}
\label{sec:application}
We survey eight application domains where the AI4SG literature has made significant progress. 
In addition to introducing the existing work, we attempt to address a broader question in the AI4SG research: what is a concrete social good problem which AI can help to address.

We propose to use three conceptual methods to systematically think about this question: in terms of the topic structure of the domain, the scope of the problem with respect to the agents involved, and functionality of the AI intervention. First and foremost, one can adopt the domain's perspective by tracing the supply chain and identify pain points along the way, thereby determining the topics of social good problems worth investigating in this domain. Using agriculture as an example, we consider two sequential topics: crop management and the agricultural produce market. One can further break down crop management into the following steps (sub-topics): crop planning, crop maintenance, crop yield prediction, and crop disease mitigation. In this way, we can identify the topics on AI for agriculture.

As the second approach, we propose a uniform structured AEC (agent -- environment -- community) framework to classify the current efforts in all AI4SG application domains based on the scope of the problem with respect to the agents involved, and to structure thoughts on identifying new problems to work on.

An agent is the smallest effect unit that has a direct interest in the output of the AI solution, e.g. a patient with some disease, a driver traveling from location A to location B, a farmer growing certain crop. 
A problem is in the ``agent'' category if we can discuss all the factors and the solution of the problem restricted to a single agent. Problems in this category include choosing the right crop for a farmer to grow and disease diagnosis based on X-ray images.

A problem is in the ``environment'' category if the problem conceptually involves multiple agents, and those agents are considered as a population without interactions and/or heterogeneity. For example, when predicting the traffic flow, typically one does not consider how the drivers interact and instead studies the system-level behavior.

A problem is in the ``community'' category if the problem conceptually involves multiple heterogeneous agents with explicit consideration of their interactions. For example, when designing the pricing mechanism of a ride-sharing platform using AI, we can take into account the drivers and riders' different incentives and preferences and their interactions.

A specific research topic might have multiple types of problems. Citizen science is scientific research conducted in part by nonprofessional (citizen) scientists who, for example, contribute data and observation to the research effort. Correcting the bias in citizen science data might belong to ``environment''. Incentivizing citizen scientists to collect better data might belong to ``community''.

The third approach is to consider the functionality of the AI intervention, and we propose to use the DPP (descriptive -- predictive -- prescriptive) framework. Descriptive analysis aims to provide insight into the current and previous phenomenon, like using animal images to get an animal census in a conservation area. Predictive analysis focuses on predicting what might happen in the future, like predicting whether a patient will be readmitted based on electronic health records. Prescriptive analysis emphasizes providing advice on what to do next, like how to allocate security resources in an airport.

In the sequel, we discuss one application domain in each subsection. For each domain, we first provide a table listing a few example problems by the AEC and DPP frameworks. Then, we discuss the current research efforts bundled by the topic structure. 
Towards the end of each subsection, we include a case study of a representative AI4SG project. We select this project from all the AI4SG projects in this domain that have been deployed/piloted in the real world. We analyze this project using a framework adapted from NetHope's AI for good guide~\cite{manual:2019:Nethope:ai4g}. For each project, we introduce the target problem, the previous practice (before the AI intervention) and explain why AI could help. After providing an overview of the AI intervention, we identify the key data and resources required for such a project. In addition, we discuss the current status of the project, especially after the paper publication. We also analyze in retrospect what part of the AI solution worked and what part did not work.
Finally, we are aware that the literature we surveyed does not contain all the work on this topic. Therefore, for each application domain, we point the interested reader to surveys of the AI methods used in the domain which are published in the application domain's research community, where applicable.

\subsection{AI for Agriculture}
\label{sec:agriculture}
\begin{table}[h]
    \centering
    \begin{tabular}{|c|c|C{10cm}|}
    \hline
    \multirow{3}{*}{Agent} & Descriptive & Crop disease diagnosis \cite{AAAI:2011:Quinn:disease}  \\
    \cline{2-3}
     &  Predictive & Crop yield prediction~\cite{AAAI:2017:You:cropyield} \\
    \cline{2-3}
     &  Prescriptive & Crop choice recommendation~\cite{AAAI:2008:VonLucken:soil} \\
    \hline
    \multirow{3}{*}{Environment} & Descriptive & Crop growth modeling (Modeling crop disease~\cite{AAAI:2011:Quinn:disease})  \\
    \cline{2-3}
     &  Predictive & \\
    \cline{2-3}
     &  Prescriptive & Growing conditions prediction (Drought prediction \cite{AAAI:2012:Kersting:drought}) \\
    \hline
    \multirow{3}{*}{Community} & Descriptive &  \\
    \cline{2-3}
     &  Predictive & Produce price forecasting~\cite{ICTD:2019:Ma:produceprice} \\
    \cline{2-3}
     &  Prescriptive & Alternate market designs (Electronic agriculture marketplace~\cite{COMPASS:2018:Newman:market}) \\
    \hline
    \end{tabular}
    \caption{AEC and DPP categorization of research on AI for agriculture.}
    \label{tab:agriculture}
\end{table}

The United Nations estimates that the world needs to feed two billion additional people by 2050~\cite{UN:2015:UN:population}, which presents a significant challenge to the global agriculture capacity.
Meanwhile, currently 45\% of the world's population lives in rural areas and 26.7\% sustains their living through agriculture~\cite{NA:2018:FAO:agriculture}.
Thus, to improve agriculture is to address both today's and tomorrow's problems.
AI has lots of potential in this domain and we believe that much more work could be done. 
Table~\ref{tab:agriculture} shows some example works on AI for agriculture and we classify them under the AEC and DPP structure. In this subsection, we follow the agriculture supply chain and group the existing works into two topics, crop management and agricultural marketplace.

\subsubsection{Crop management}
The most intuitive and fundamental goal of AI for agriculture is to improve crop management and boost crop yield. The progress and commercialization of UAVs, satellite images, and IoT have rendered data that are unprecedented in both quantity and precision. These data sources enable AI to boost the efficiency of farming, promoting the so-called ``precision agriculture''. 
With them, it is possible to address the various aspects of crop management, such as crop planning, maintenance, yield prediction, mitigating crop disease, and agricultural information gathering.

\paragraph{Crop planning} Deciding on what crop to grow and when to grow could be seen as the first step of crop management. \cite{AAAI:2008:VonLucken:soil} uses multi-objective evolutionary algorithms to determine, based on the soil information, the optimal crop to grow for sustainable land use. 
The optimization criteria include the costs of fertilizing and liming, cultivation, and the expectation and variation of total return.
Other works explore selecting the best time to sow based on additional factors such as the weather information. A pilot study in India instructed farmers to delay planting by three weeks using predictive models based on climate and weather data and increased the yield by 30\%~\cite{NEWS:2017:ICRISAT:whentosow}. 

\paragraph{Crop maintenance}
Second, crop maintenance is critical to the yield. There are many aspects where AI can help. On the one hand, advanced sensing technologies show much promise for AI. On the other hand, these headline technologies are still far from reaching an average farmer. AI can also help to leverage existing tools and resources for farmer's use. For example, crop evapotranspiration information is useful for optimizing irrigation plans, yet its accurate measurement is limited due to the sparse evapotranspiration network. \cite{IJCAI:2013:Holman:irrigation} propose using the Gaussian Process to estimate this information with data from ordinary weather stations, and the approach has been demonstrated to work well in Texas.

\paragraph{Crop yield prediction}
Third, being able to predict the crop yield can often help farmers plan for the coming year and help the government determine agricultural policy. While lots of work have developed the key features to make this prediction, these features are often costly to compute in practice. Remote sensing products, specifically satellite images, provide a cheap alternative. \cite{AAAI:2017:You:cropyield} not only use deep learning models to predict the crop yield from satellite images but also design a compact representation of the images to address the scarcity of labeled data and use a Gaussian Process model to address the spatial-temporal variation.

\paragraph{Mitigating crop disease}
Crop disease is a major cause of poverty in developing countries. Quinn et al.~\cite{AAAI:2011:Quinn:disease} propose several approaches to address problems related to crop disease, including using Gaussian Process ordinal regression to estimate disease distribution, using this model for survey planning, and using CV techniques including scale-invariant feature transform on photos for crop disease diagnosis. A mobile app based on~\cite{AAAI:2011:Quinn:disease} and its follow-up works is now available~\cite{website:2016:mcrops}.

\paragraph{Agricultural information gathering}
At a macro level, it is also valuable to gather data on which crops are grown where.
\cite{AAAI:2017:You:cropyield} makes use of deep Gaussian processes to infer agriculture information from satellite imagery. \cite{COMPASS:2019:Jain:aerial} enables farmers to monitor their crops at low cost via aerial imaging, which relies on CV and path-planning algorithms.

\subsubsection{Agriculture produce market}
Aside from dealing with crops in the field, the agriculture market is also vital to a farmer's life. AI techniques such as time series analysis and algorithmic mechanism design have much potential in this domain. Being able to predict the price movement can greatly benefit the farmers, as the fluctuation of produce price directly determines a farmer's income. An agricultural market, especially in developing countries, is different from, say, a stock market in its lack of information and infrastructure. \cite{ICTD:2019:Ma:produceprice} use the nearest neighbor model to predict the agricultural market price in India. 

However, even if we can predict the produce prices accurately, the farmers may still face difficulties on the market. These markets have limited information flow and high transaction costs and pose a significant requirement to the mobility of the farmers. Thus, other works have explored setting up a better market to connect the farmers. \cite{COMPASS:2018:Newman:market} used over five years to set up a text message-based agricultural market in Uganda. We discuss this work in more detail in the following case study.

\subsubsection{Case study: Kudu -- an agricultural market in Uganda}

\paragraph{Target problem}
Farmers in Uganda often lack adequate information about the agricultural market in the country, which puts themselves at a disadvantage when they sell their produces. Newman et al. attempt to set up, Kudu, an electronic agricultural market to facilitate the information flow~\cite{COMPASS:2018:Newman:market}. The problem fits into ``community -- prescriptive'' in the AEC -- DPP framework.

\paragraph{Previous practice}
There are markets in both cities and rural areas, but they are quite different in terms of the scale, price, and participants. These markets often feature inconsistent availability of produce and arbitrage opportunities. For example, the country has lots of tomato growers but occasionally one cannot find any tomato at the roadside markets. One can do temporal and spatial arbitrage without much difficulty.
A main reason is that individual farmers in the rural areas often do not have sufficient information about the national market, and as a result the farmers also miss lots of opportunities.

\paragraph{Why is AI needed}
The primary problem is to make it easier for individual farmers to get access to the national market. Uganda has excellent cell phone coverage which makes an electronic marketplace possible. However, it is too much cognitive burden for a human to match the sellers and buyers, while AI could automate the process and ensure that the new marketplace functions properly. In addition, the electronic marketplace has great value in itself, because it saves the farmers from point-to-point face-to-face communication and makes the transactions more efficient.

\paragraph{Intervention overview}
The authors proposed and implemented a text message-based agricultural market using mechanism design. Both buyers and sellers can post to the market. There is a central clearinghouse and the authors have attempted an AI-based clearing mechanism which runs a maximum weight matching algorithm as well as a hybrid system that combines AI-based and manual clearing. Once a bid and ask are matched, the bidder and the seller will travel to meet in person and the transaction happens offline.

\paragraph{Data used} 
To propose a match, the authors used Uganda's road network data to incorporate the transportation cost into consideration, which can be readily obtained online. Other values core to the market are not data-driven and are estimated based on heuristics. However, the local farmers are reluctant to adopt the system unless the authors also provide the average market price. These data are easy to obtain once the authors have collaborated with a local partner to obtain these data.

\paragraph{Resources needed}
Beyond doubt, any implementation and maintenance of a market like this need continued funding. The authors collaborated with several local agencies, who are vital to the deployment of any kind, in terms of local knowledge, social connections, local outreach, etc.

\paragraph{Deployment status}
Since its initial deployment in 2013, Kudu has attracted over 21000 users and facilitated over 850 verified transactions, with a total transaction value of over 1.9 million USD. In comparison, the agricultural share of Uganda's GDP was around 6.5 billion USD. Thus, there is no doubt that Kudu has obtained a significant scale and reach.
In the paper~\cite{COMPASS:2018:Newman:market}, it is stated that the grant funding for this project ended in March 2018. This is also the time that Kudu's website\footnote{\url{https://kudu.ug/}} stopped updating. Nevertheless, the authors report that their local partners planned to continue the operation.

\paragraph{AI in retrospect}
The core of a designed market is its market-clearing algorithm. Some of the desired properties of a market are motivated by the difficulties in their previous pilots. For example, too sparse clearing frequency drives the farmers away from the platform because people do not want to wait that long, and a continuous clearing thus might be desirable. Yet it is hard to analyze whether some other properties are effective in practice. For example, one desired property in mechanism design is to maximize the social welfare, and in this case, the total gain from trade. However, not all trades matched on the platform actually happened due to the primitive trade execution situation. Thus, maximizing welfare at clearing might not be useful in practice. In general, the work suggests that human clearing cannot be substituted by algorithmic clearing in Uganda and a hybrid approach might be more reasonable. Furthermore, the authors choose to base the market primarily on text messages, as they observed that text messages are much more accessible than, say, computer and the Internet. However, during deployment, most of the transactions happened through phone calls with human coordinators, an arguably low-tech intervention compared to what the authors intended. That said, this is one of the very few attempts from the community to build a non-trivial market from scratch. Understanding all the practical constraints introduces to the research community many valuable problems to ponder.

\subsubsection{Other works}
Outside the AI community, there have been lots of attempts to apply AI/ML techniques to address agricultural issues. 
We refer the interested reader to~\cite{MDPI:2018:Liakos:agriculture} for a survey of these works.
\subsection{AI for Education}
\label{sec:education}
\begin{table}[h]
    \centering
    \begin{tabular}{|c|c|C{10cm}|}
    \hline
    \multirow{3}{*}{Agent} & Descriptive &  Predictive evaluation (Modeling student learning state \cite{IJCAI:2019:Grover:vocabulary}) \\
    \cline{2-3}
     &  Predictive & Individual performance prediction \cite{AAAI:2018:Su:exercise} \\
    \cline{2-3}
     &  Prescriptive & Problem and solution generation \cite{IJCAI:2015:Wang:geometry} \\
    \hline
    \multirow{3}{*}{Environment} & Descriptive & Modeling student engagement \cite{AAAI:2014:Ramesh:latent}  \\
    \cline{2-3}
     &  Predictive & Student performance prediction (Identifying at-risk students \cite{AAAI:2015:He:atrisk})\\
    \cline{2-3}
     &  Prescriptive & Online education platforms \cite{IAAI:2019:Roy:prerequisite}\\
    \hline
    \multirow{3}{*}{Community} & Descriptive &  \\
    \cline{2-3}
     &  Predictive & Problem difficulty prediction \cite{AAAI:2017:Huang:difficulty} \\
    \cline{2-3}
     &  Prescriptive & Administration improvement \cite{IAAI:2013:Cambazard:timetable} \\
    \hline
    \end{tabular}
    \caption{AEC and DPP categorization of research on AI for education.}
    \label{tab:education}
\end{table}

For many, education represents and enables future growth and opportunity. Literacy in language, math, and sciences, and other essential skills empower individuals to participate in today's global economy and perhaps break cycles of poverty. However, the United Nations reports that in low-income countries, 91\% of students in primary school fail to achieve minimum reading proficiency, and 87\% fail to achieve basic math proficiency \cite{UNESCO:2017:UN:literacy}. The application of AI to this domain has the potential to provide equal opportunities to all students worldwide, thereby investing in future innovation and global development. We believe much more work could be done in this domain. We classify some sample works on AI for education in Table~\ref{tab:education} using the AEC and DPP structure. In this subsection, we introduce the existing works on both teaching a specific skill and education issues in a community environment.

\subsubsection{Students and a (virtual) teacher}
Personalized education has long been an ideal for families around the world. Although technology has permeated into many aspects of education, the fundamental ways in which the teachers teach and the students learn have not changed much. The progress of AI is now starting to bring us a little bit closer to this ideal. The first step is to develop AI which functions as an assistant to, or even a substitute for, teachers.
We discuss problem-level and curriculum-level problems, two types of tasks that AI has been used to address, which differ by the length of interaction with the student.

\paragraph{Problem-level AI for education} 
In this category, AI is developed to facilitate a concrete procedure in the student's learning process, including problem generation, problem-solving, grading, and difficulty prediction. These works necessarily focus on a specific discipline and leverage the special structure of that domain. For example, Wang and Su propose to use natural language generation aided with dimensional units to generate math exercise problems~\cite{IJCAI:2016:Wang:mathword}. They also developed automated reasoning techniques to generate problem solutions~\cite{IJCAI:2015:Wang:geometry}. Similar tasks have been attempted in logic problems~\cite{IJCAI:2013:Ahmed:naturaldeduction}. When the problem has a definitive solution, works on grading focus on giving proper credit to partial solutions. Without a definitive correct answer~\cite{IJCAI:2013:Alur:dfa}, such as essay writing, one could use natural language processing to assess the quality and provide feedback~\cite{IAAI:2019:Zhang:erevise}.
Computer programming is a key area that has seen lots of development in terms of generating pattern programs~\cite{IJCAI:2018:So:synthesizing} and program error checking~\cite{AAAI:2019:Gupta:syntactic,AAAI:2017:Gupta:deepfix,AAAI:2019:Wu:rubric}.
One level above, there is also work that explore generating an exam with the right combination of problems against a student population~\cite{IJCAI:2015:Geiger:exam}, and predicting the difficulty of the problems~\cite{AAAI:2017:Huang:difficulty}.

\paragraph{Curriculum-level AI for education}
Intuitively, helping a student master a curriculum is intrinsically different from grading a student's essay since it takes time and repeated effort to master a skill which could have many building blocks.
The tasks in this category typically feature repeated interaction with the student. Intelligent tutoring systems (ITS), which use AI techniques to provide step-by-step tutoring, monitor student's progress, and offer instant feedback, are not a new concept~\cite{BOOK:1988:Burns:its}. Recent works integrate a data-driven approach like graphical models~\cite{IAAI:2011:Green:dbn,IJCAI:2019:Grover:vocabulary}, factorization machines~\cite{AAAI:2019:Vie:knowledgetracing}, or recurrent neural networks~\cite{AAAI:2018:Su:exercise} into ITS to model the student's learning state. In addition, Valentine et al. employ sketch recognition techniques to assist instructors in homework assignments~\cite{IAAI:2012:Valentine:mechanix}. 

Researchers also tried to develop AI technologies to assist student learning beyond the well-established ITS framework. For example, Park et al. developed a social bot companion for early literacy education~\cite{AAAI:2019:Park:social}. They use reinforcement learning to find an optimal policy to engage with each given student. Zhang et al. use logic programming to facilitate the pedagogy of computer programming in the STEM subjects~\cite{EAAI:2019:Zhang:logicprogramming}.

\subsubsection{Students in an educational environment}
In addition to the specific setting of learning a particular skill, the environment in which education happens is also important. On this front, we cover two topics. First, we consider the online education platforms which are gaining more presence. Second, we introduce the work on student's long-term success in an education system like high school or college.
\paragraph{Education on online education platforms}
The advent of online education platforms such as Coursera and EdX brings about a new dimension of education that differs from the abovementioned works. Specifically, it has a social aspect: students interact, collaborate, and social with millions of anonymous learners through the Internet. Strictly speaking, this is not new to education, as most people have classmates throughout their schooling career. Yet, the public, structured, and logged interactions allow the AI researchers to dive deeper. 
For example, He et al. use the students' online discussions in MOOCs to measure student attributes such as attitude, ability, and intelligence~\cite{AAAI:2016:He:mooc}. In a classroom environment, this may only be assessed by assignments and exams. A serious problem in MOOCs is student dropout. Recent works combine graphical models with first-order logic to analyze students' engagement patterns~\cite{AAAI:2014:Ramesh:latent}, and use logistic regression to help decide intervention on students at risk of dropout~\cite{AAAI:2015:He:atrisk}. Roy et al. study the problem of inferring pre-requisites for MOOCs~\cite{IAAI:2019:Roy:prerequisite}. 

Traditional classrooms have also adopted some features in MOOCs such as online discussion forums. Kim and Shaw investigated how to organize and retrieve past discussions to help the current students learn~\cite{IAAI:2009:Kim:pedabot}.

\paragraph{Student success in the education system}
In addition to helping a student master a skill or a course, AI is also a useful tool to study the student's overall success (or failure) over an education stage. Researchers have shown that lots of factors could be used to predict education outcomes, such as previous academic performance, family background, and even Internet usage~\cite{AAAI:2019:Trakunphutthirak:thai}. Xu et al. use an ensemble learning method to predict students' college performance continuously as the student proceeds in the program~\cite{AAAI:2017:Xu:college}. In such predictive analysis, typical metrics like precision and recall may not be the best performance indicator of the model. Lakkaraju et al. propose a few evaluation metrics in the context of preventing high school dropouts~\cite{KDD:2015:Lakkaraju:highschool}.

\subsubsection{Case study: scaffolding student online discussions using past
discussions}

\paragraph{Target problem}
At many universities, instructors use online discussion platforms as an auxiliary tool to facilitate collaborative learning among students. However, these platforms are often either underutilized or fail to achieve their potential of fostering peer-to-peer interactions (only delivering the student-to-teacher interactions). Kim and Shaw attempted to facilitate student interaction using past discussions~\cite{IAAI:2009:Kim:pedabot,AIR:2014:Kim:pedabot}. This problem fits into ``environment -- descriptive'' in the AEC -- DPP framework.

\paragraph{Why is AI needed}
At the time of their paper~\cite{IAAI:2009:Kim:pedabot}, online discussion platforms had very few features that directly foster their usage and usability. The authors hypothesized that, if students, when they post a question, are presented with related discussions on the same topic from the previous iterations of the course, such information can prompt them to reflect and use the discussion platform more actively. AI is required to achieve this information retrieval task.

\paragraph{Intervention overview}
The authors use existing information retrieval techniques like tf-idf to compute the similarity between the student's new post and previous posts on the platform. The most relevant previous posts are then selected to display to the student. The authors also attempted topic modeling techniques which accommodate multi-word domain terms and acronyms as well as latent semantic analysis transformations yet they reported that these techniques did not yield satisfactory performance. The authors have deployed the tool in three computer science courses at the University of Southern California.

\paragraph{Data used} 
The AI intervention uses message corpus from the discussion platform during the historical offerings of the same course, as well as some text from the course textbook. The AI intervention directly selects from these data as its output. Intuitively, in the university environment, obtaining these data should not pose significant difficulty.

\paragraph{Resources needed}
Evidently as an educational tool in a university environment, the research received the support from the course instructor and the university. As a live system during a course offering, the researchers needed to pay close attention throughout. For example, some students mentioned that the names were not attached to the retrieved messages so they could not identify the instructor's response from students' posts. This was fixed during the deployment.

\paragraph{Deployment status}
The AI intervention was deployed in a computer science course for three semesters at the University of Southern California. It reached a total of 469 students and witnessed over 2000 messages posted on the discussion platform. In some of the semesters deployed, the tool facilitated significantly longer threads of discussion compared to not using the tool.
Online discussion platforms have evolved quickly since this project. Today's similar platforms have many additional tools to facilitate collaborative learning, yet the very function that this project explored is missing from today's mainstream platforms.

\paragraph{AI in retrospect}
In terms of the original motivation of the research, the AI intervention did lead to an increased usage of discussion platforms in some experiments. However, the relevance of Pedabot's retrieved discussions could be improved, as commented by several students who used the system. This perhaps hindered the further deployment of the research.

\subsubsection{Other works}
The AI for education community has studied the potential of AI in education for a long time and has developed a rich literature. The interested readers could refer to the International Conference on Artificial Intelligence in Education\footnote{\url{https://iaied.org/conferences}} (AIED) for more research in this community. For a survey of intelligent tutoring systems, please refer to~\cite{BOOK:1988:Burns:its,IJCA:2019:Alkhatlan:itssurvey}.

\subsection{AI for Environmental Sustainability}
\label{sec:sustainability}

\begin{table}[h]
    \centering
    \begin{tabular}{|c|c|C{10cm}|}
    \hline
    \multirow{3}{*}{Agent} & Descriptive & Energy usage modeling \cite{AAAI:2015:Li:energyusage}  \\
    \cline{2-3}
     &  Predictive & Energy usage predictions (Forecasting Multi-Appliance Usage \cite{IJCAI:2013:Truong:multiappliance}) \\
    \cline{2-3}
     &  Prescriptive & Smart homes (Sustainable temperature tuning \cite{IJCAI:2019:Fiducioso:roomtemp})\\
    \hline
    \multirow{3}{*}{Environment} & Descriptive & Biodiversity information gathering (Acoustic monitoring of the African forest elephant \cite{AAAI:2019:Bjorck:elephant}) \\
    \cline{2-3}
     &  Predictive & Climate prediction (Predicting temperature volatility \cite{AAAI:2019:Khodadadi:volatility}) \\
    \cline{2-3}
     &  Prescriptive & Wildlife protection (Defense against poaching and illegal fishing \cite{IJCAI:2015:Fang:gsg})\\
    \hline
    \multirow{3}{*}{Community} & Descriptive & Complex ecosystem modeling (Modeling interaction between species \cite{AAAI:2012:Martinez:sustaining}) \\
    \cline{2-3}
     &  Predictive & Smart grid prediction (Forecasting clear skies for power grid \cite{IJCAI:2017:Palani:skies})  \\
    \cline{2-3}
     &  Prescriptive &  Energy distribution (Smart grid energy distribution marketplace \cite{AAMAS:2015:Cerquides:energy})\\
    \hline
    \end{tabular}
    \caption{AEC and DPP categorization of research on AI for environmental sustainability.}
    \label{tab:environment}
\end{table}

The long-lasting and alarmingly novel effects of human activity on Earth have been known for decades: global temperatures have risen to unprecedented levels, amounts of ice have diminished, and species the world over have been driven to extinction. Immediate action must be taken lest we allow the planet to become uninhabitable for ourselves and other species, and AI can help to take such action in an informed and efficient manner. In fact, this is one of the most popular domains studied by the AI4SG community. In Table~\ref{tab:environment}, we include some sample works on AI for environmental sustainability, classifying them under the AEC and DPP structure. We introduce the existing works on three main topics in environmental sustainability: climate, resources, and biodiversity.

\subsubsection{Climate monitoring and prediction}
Slowing the progress of climate change and/or mitigating its effects are at the forefront of AI for sustainability goals. 
A basic step is to use AI to monitor and predict climate and weather conditions. There is a rich literature on predicting various climate variables, such as temperature and precipitation. A few works approach this task as (some variants of) a LASSO problem~\cite{AAAI:2019:Khodadadi:volatility,AAAI:2019:He:lasso,AAAI:2016:Chatterjee:greatlake}, which is a linear regression model with $l^1$-regularization. Others leverage the connection between different climate variables at neighboring locations to provide combined predictions~\cite{AAAI:2017:Gonccalves:multitask,AAAI:2012:McQuade:neighborhood}. Gaussian process is another commonly used tool for climate monitoring and prediction~\cite{AAAI:2012:Garg:gp,AAAI:2017:Vu:electricitybill}. Rather than using typical environment predictors, Vu and Parker extract urban microclimates from electricity consumption data~\cite{AAAI:2017:Vu:electricitybill}. 

The degradation of air quality is a major threat in a number of the world's developing regions. It has also received rising attention from the AI community. Existing works use both Gaussian process~\cite{AAAI:2015:Guizilini:nonparametric} and deep learning methods~\cite{IJCAI:2018:Liang:geoman,AAAI:2018:Cheng:airquality} to monitor and predict air quality.
Community sensing is an attractive alternative to the traditional sensor-based measurement and prediction~\cite{AAMAS:2015:Zenonos:pollution,AAAI:2018:Bian:cswa}.
For example, Zenonos et al. propose a participatory sensing network based on local greedy search which assigns each human participant to air pollution measurements that need to be taken~\cite{AAMAS:2015:Zenonos:pollution}.

\subsubsection{Resource conservation}
Conservation of natural resources such as water, land, air, and energy is crucial to environmental sustainability, ensuring that those from all parts of the planet and future generations have the resources needed to survive. 
In particular, energy conservation has received the most attention and in the sequel we introduce the relevant works on smart home, smart grid, and electric vehicles. We also describe the works on other resource conservation towards the end.

\paragraph{Smart home energy management} 
Firstly, modern homes are often equipped with intelligent, connected devices which aid homeowners in fine-tuning their households towards efficient utility usage. For example, a smart dishwasher might be able to record how much energy it has consumed in a month, and a smart laundry unit might turn itself on during off-peak hours. Various novel applications have been shown to effectively model and predict energy usage among households~\cite{IJCAI:2015:Auffenberg:thermalcomfort}. Li and Zha model energy usage behavior of singular appliances within a household using Hawkes processes~\cite{AAAI:2015:Li:energyusage}. Truong et al. similarly model domestic usage of multiple appliances, taking into account habitual routines and interdependence usage of different appliances~\cite{IJCAI:2013:Truong:multiappliance}. Through Safe Contextual Bayesian Optimization, Fiducioso et al. optimize the temperature control in rooms to minimize energy consumption while maintaining inhabitant comfort and safety~\cite{IJCAI:2019:Fiducioso:roomtemp}.
Based on these results, AI applications can then make recommendations for optimizing energy usage within singular buildings or households in an efficient manner.

\paragraph{Smart grid} 
Smart home energy management plays an important role in the bigger picture of optimizing energy distribution through a city or state at the electricity grid level.
The smart grid, as the vision of future electricity grid, features a two-way interaction between the producers and consumers in a sustainable, economic and secure manner~\cite{NA:2012:Smart:smartgrid}. For example, a smart grid can more reliably handle distributed (consumer) energy generation, while the traditional electricity grid was designed for one-way electricity flow. On the consumer side, energy consumption fluctuates greatly, and thus it is valuable to predict future consumption in order to conserve energy and avoid incidents.
A few works predict the power consumption on a power grid~\cite{IJCAI:2018:Chen:neucast,AAAI:2019:Salem:intrahour}. In particular, in the work of Salem et al.~\cite{AAAI:2019:Salem:intrahour}, a prototype was being developed to be tested in the Norwegian state grid. 
On the producer side, other works formulate the energy supply restoration problem as an optimization problem~\cite{AAMAS:2015:Agrawal:decentralized,IJCAI:2013:Thiebaux:planning}.
Finally, it is natural to take a multi-agent system perspective on the smart grid~\cite{AAMAS:2011:Chalkiadakis:cooperative}.
Cerquides et al.~\cite{AAMAS:2015:Cerquides:energy}, for instance, proposes a model for a novel decentralized marketplace to trade and distribute energy in an electricity network. The market's allocation rule uses dynamic programming techniques and has been shown in empirical results that it outperforms benchmark optimization problem solvers in this setting. 

\paragraph{Electric vehicles}
As electric/hybrid vehicles make their way into more people's lives, they also received special attention from the AI community. For example, Vogel et al. studied improving hybrid vehicle fuel efficiency by predicting the driver's route choice using inverse reinforcement learning~\cite{AAAI:2012:Vogel:irl}. Lots of works have studied the placement of electric vehicle charging stations~\cite{IJCAI:2015:Xiong:evplacement}, and the charging mechanisms as the current power grid is not designed for large-scale charging load~\cite{IJCAI:2015:Hayakawa:ev,AAMAS:2011:Gerding:ev,AAMAS:2015:Valogianni:ev}. Kahlen and Ketter consider the two-way energy transaction between electric vehicles and electricity grid~\cite{AAAI:2015:Kahlen:carcharge}.

\paragraph{Other resource conservation}
Aside from energy, water conservation has been studied from multiple angles, such as the management of water resource systems~\cite{AAMAS:2015:Amigoni:water}, water consumption disaggregation~\cite{IJCAI:2013:Dong:disaggregation}, water-efficient landscape design using plant facilitation~\cite{AAAI:2011:Hoenigman:facilitation}. Kanters et al. use online planning to optimize pumping station control in the Netherlands~\cite{AAAI:2016:Kanters:pumping}. It is also important to note that resource conservation is usually intertwined with other environmental and social indicators, and thus researchers might find themselves solving a multi-objective optimization problem. For example, Wu et al. developed dynamic programming and mixed integer programming algorithms for approximating the Pareto frontier in the problem of hydropower dam placement in the Amazon basin~\cite{AAAI:2018:Wu:dam}.

\subsubsection{Biodiversity}

Biological diversity provides ecosystems resilience and is integral to their survival and recovery (after incidents like natural disasters). Because humans are ultimately part of the same ecosystem as all plants and animal species on a macroscopic scale, biodiversity benefits our quality of life and reinforces our survival as well.

\paragraph{Biodiversity monitoring}
There are a few ways to gather information about the distribution of animals and plants, which is fundamental to conservation efforts. Citizen science projects involve the general public in the process by having people report their observations. Based on the eBird project which gathers observations from the bird watchers, Hutchinson et al. model the birds' distribution as a classification problem with class-conditional noise~\cite{AAAI:2017:Hutchinson:noise}. The Wildbook project uses computer vision on photos provided by citizen scientists to identify each individual animal, and thereby providing an animal census~\cite{Bloomberg:2017:Bergerwolf:wildbook}.
A few works use acoustic information to detect the presence of certain animals such as cicadas~\cite{IJCAI:2013:Zilli:cicada}, elephants~\cite{AAAI:2019:Bjorck:elephant}, and birds~\cite{IJCAI:2015:Ruiz:birdsong}. These works leverage different techniques, including hidden Markov models~\cite{IJCAI:2013:Zilli:cicada}, deep neural networks~\cite{AAAI:2019:Bjorck:elephant}, and unsupervised segmentation of birdsong recordings and multiple instance classification~\cite{IJCAI:2015:Ruiz:birdsong}. All above works output information about a single species. There are also a few works that try to characterize the distributions of multiple species simultaneously~\cite{IJCAI:2017:Chen:embedding,AAAI:2018:Tang:multientity}.

\paragraph{Habitat planning}
Recognizing and managing habitats are important aspects of conservation. Zheng et al. use generative adversarial networks (GAN) to map animal habitats in the Qinghai Lake breeding ground in China based on remote sensing imagery~\cite{AAAI:2019:Zheng:habitat}.
A number of works have focused on the habitat planning problems, such as using constraint programming for the partitioning of conservation areas~\cite{IJCAI:2019:Justeau-Allaire:partition} and for the habitat restoration effort~\cite{AAAI:2016:Urli:restoration}, and using submodular optimization for sequential resource allocation to different habitats~\cite{AAAI:2011:Golovin:conservation}.
Faced with the increasing fragmentation of habitats, a few researchers consider the connectivity of habitats in conservation efforts~\cite{AAAI:2017:Xue:connectivity,AAAI:2017:Wu:tree,AAAI:2013:LeBras:multispecies}. These works typically formulate a mathematical program and solve it using techniques from MIP and heuristic search.

\paragraph{Species conservation} Every natural ecosystem is delicately balanced by the coexisting species of plants and animals which live in it. The endangerment and extinction of species in the wild, often due to human activity such as poaching and deforestation, disrupts this homeostasis. 
Game theory is a natural tool to model the adversarial interaction between the conservation agency and the poachers. A large body of work on the Stackelberg security games has studied the game-theoretic model as well as been deployed in the real world. The green security game addresses the domain of wildlife conservation~\cite{IJCAI:2015:Fang:gsg,AAMAS:2016:Nguyen:capture}. Its algorithm, PAWS, has been deployed in a number of conservation sites~\cite{IAAI:2016:Fang:gsg}. There is also literature on security games being applied to protecting forests~\cite{AAAI:2012:Johnson:forest,AAAI:2016:McCarthy:logging}, fisheries~\cite{IAAI:2014:Haskell:fisheries}, and coral reefs~\cite{IJCAI:2016:Yin:coral}.

\paragraph{Invasive species management} Invasive species, or species which are introduced to a new environment where they are not native, can similarly cause ecological harm and threaten the balance of that environment's natural ecosystem. Gupta et al.~\cite{IJCAI:2018:Gupta:hawkes} intervene with the spread of invasive species by first modeling the spread using Hawkes processes, then deriving and optimizing closed-form expressions as an integer program. Peron et al. use a Markov decision process (MDP) to compute the optimal intervention plan to stop the invasive species from spreading to some location~\cite{AAAI:2017:Peron:momdp}. Another work takes the perspective from using biological control agents as a countermeasure, where Spencer considers a graph vaccination problem~\cite{AAAI:2012:Spencer:robustcut} and then solves it using submodular maximization algorithms.

\subsubsection{Case study: protection assistant for wildlife security}
\label{sec:paws}
\paragraph{Problem to be solved}
Wildlife poaching is a great threat to the biodiversity of our planet. Because of the high profit of poaching, the poachers have become increasingly sophisticated to outsmart the patrollers. Thus, it is important to design patrol routes in a strategic way. In a line of works~\cite{AAMAS:2014:Yang:firstpaws,IAAI:2016:Fang:gsg,ECML:2017:Gholami:cft,ICDE:2020:Gholami:fieldtest}, researchers have developed and deployed a suite of game-theoretic tool PAWS to combat poaching. This problem falls into ``community -- prescriptive'' under the AEC -- DPP framework.

\paragraph{Why is AI needed}
Patrollers in wildlife conservation areas have lots of experience conducting patrols. They design the patrol routes based on their knowledge and experience in the area. However, since the poachers are highly strategic in evading the patrollers, these patrol routes are very susceptible to gaming. In addition, while experience can help the patrollers, it may also keep the patrollers from going to underexplored areas where the poachings might be frequent. A game-theoretic planner for patrol routes can be useful in addressing these issues.

\paragraph{Intervention overview}
The authors propose PAWS (Protection Assistant for Wildlife Security), which is based on several existing algorithms for the Stackelberg security games.\footnote{For more discussion on Stackelberg security games, see Section~\ref{sec:securitygame}.} PAWS models a two-player zero-sum game between an attacker (the poacher) and a defender (the patroller). The authors use ML to learn poacher's behavior patterns from historical data. The ML methods used have gone through several stages, from the subjective utility quantal response model~\cite{AAMAS:2014:Yang:firstpaws}, to dynamic Bayesian networks~\cite{AAMAS:2016:Nguyen:capture}, to a variant of decision tree ensembles~\cite{AAMAS:2017:Kar:decisiontree}, to a hybrid model of decision trees and Markov random fields~\cite{ECML:2017:Gholami:cft}, and to Gaussian Processes~\cite{ICDE:2020:Gholami:fieldtest}. Some of the works also leverage domain expert's input, using them to augment the dataset~\cite{COMPASS:2018:Gurumurthy:poaching}. Solving for the equilibrium of this game through mathematical programming gives an optimal patrol strategy for the patroller~\cite{GameSec:2017:Xu:blackbox}. Further developments provide coordinated patrol plans for both human patrollers and conservation unmanned aerial vehicles (UAVs)~\cite{AAAI:2020:Bondi:uav}, and use online learning to design patrols to trade off exploitation and exploration~\cite{AAMAS:2019:Gholami:minion}. PAWS takes the topographic information into account, so that the patrol routes that PAWS outputs are compatible with the terrain.

\paragraph{Data used} 
PAWS uses the animal activity data to estimate the animal density which plays a role in determining the payoff of each patrol route. PAWS also uses the poacher activity data to aid the poacher modeling. All these data and previous patrol tracks are obtained from the collaborating conservation agency. To consider the terrain and elevation information, PAWS also uses the topographical data.

\paragraph{Resources needed}
The development of PAWS spans multiple years and multiple publications, and multiple researchers' human resources. All the work which add to PAWS the bounded rationality model predictive analysis, and the incorporation of UAV planning contributed to a research product that has the potential to wide deployment today. Throughout this work, the collaboration with conservation agencies is key to identifying new research problems and putting the results in the field.

\paragraph{Deployment status}
In the series of work mentioned above, PAWS has been field tested in multiple conservation sites in Uganda, Cambodia, Malaysia, and China. The authors claim that PAWS proved to be effective in all these deployments, often leading the patrollers to patrol routes never used before but discovering poacher activities on those routes. 
The initial effort for commercialization did not prove successful.\footnote{AMOURWAY (now named Avata AI). \url{https://www.avata.ai/}} Nevertheless, the development of PAWS did not stop with these papers. In 2019, PAWS started the partnership with SMART\footnote{\url{https://smartconservationtools.org/}} (a spatial monitoring and reporting tool commonly used by wildlife rangers) and Microsoft. This partnership will allow PAWS to be used at over 800 wildlife conservation sites worldwide in the near future.

\paragraph{AI in retrospect}
The initial version of PAWS did not take the terrain and elevation into account when designing patrol routes. The collaborating conservation agency invited the authors to go for a patrol themselves to experience in person the impact of terrain and elevation. The authors then successfully addressed this issue by explicitly using ``feasible routes'' in the security game model.

\subsubsection{Other works}

A recent article by Gomes et al. contains an overview of the research efforts in computational sustainability~\cite{CACM:2019:Gomes:compsust}, which is more general than the environmental sustainability we covered in this section and has some overlapping with our other sections.
\subsection{AI for Healthcare}
\label{sec:healthcare}
\begin{table}[h]
    \centering
    \begin{tabular}{|c|c|C{10cm}|}
    \hline
    \multirow{3}{*}{Agent} & Descriptive & Disease diagnosis (Automated glaucoma diagnosis~\cite{AAAI:2019:Liao:glaucoma})  \\
    \cline{2-3}
     &  Predictive & Disease development predictions (Early prediction of diabetes complications~\cite{AAAI:Liu:2018:diabetes}) \\
    \cline{2-3}
     &  Prescriptive & Treatment recommendation (Adaptive epilepsy treatment~\cite{IAAI:2008:Guez:epilepsy}) \\
    \hline
    \multirow{3}{*}{Environment} & Descriptive & Drug development (Drug similarity integration~\cite{IJCAI:2018:Ma:drug})  \\
    \cline{2-3}
     &  Predictive & Epidemic prediction (Surveillance on epidemic dynamics~\cite{AAAI:2018:Pei:epidemic}) \\
    \cline{2-3}
     &  Prescriptive & Preventing spread of illness (Preventing foodborne illness~\cite{AAAI:2016:Sadilek:nemesis}) \\
    \hline
    \multirow{3}{*}{Community} & Descriptive & Patient subtyping \cite{KDD:2017:Baytas:subtyping}   \\
    \cline{2-3}
     &  Predictive & Predictive phenotyping~\cite{IJCAI:2019:Fu:phenotyping} \\
    \cline{2-3}
     &  Prescriptive & Transplant matching \cite{AAAI:2016:Yoon:organtransplant} \\
    \hline
    \end{tabular}
    \caption{AEC and DPP Categorization of research on AI for healthcare.}
    \label{tab:healthcare}
\end{table}

According to the World Health Organization, the state of global health has been generally on the rise. However, there remain areas of improvement, e.g. infant mortality or global healthcare, for which further developments will enhance human well-being. AI has incredible potential in the medical domain, and we believe that a great amount of work has already been done. We showcase some example works in Table~\ref{tab:healthcare}, categorized with the AEC and DPP structure. In this subsection we discuss how AI has augmented the many sides of both clinical health and public health.

\subsubsection{Clinical health}
The majority of research efforts in AI for healthcare are focused on clinical health. The wide availability of electronic health records (EHR) and medical images have directly facilitated the AI research in disease diagnosis, clinical treatment, and clinical prediction. We also survey a few other specialty areas to which AI has made significant contributions, including hospital management, organ transplant matching, drug development, mental health, and well-being.

\paragraph{Disease diagnosis}
Disease diagnosis is the topic in healthcare that has seen the most applications of AI. Recent advances in the image processing capabilities of deep learning methods are readily applicable to medical images such as radiology results. Researchers have used X-ray, MRI, and CT images for the diagnosis of brain disease~\cite{AAAI:2019:Wang:connectivity,AAAI:2015:Xin:smri} like Alzheimer's Disease~\cite{AAAI:2015:Zhe:alzheimer,IJCAI:2017:Xu:alzheimer,KDD:2013:Xiang:alzheimer}, prostate disease~\cite{AAAI:2017:Yu:prostate,AAAI:2017:Yang:prostate}, and various types of cancer~\cite{AAAI:2016:Chen:mitosis}. Several new problems accompany these advances. Irvin et al. propose a chest radiograph dataset with expert comparison~\cite{AAAI:2019:Irvin:chexpert}, since despite the seemingly abundant data, well-curated realistic datasets are still a craving of the researchers. Chen et al. address the domain adaptation problem which is common in medical treatment, for example, when a classifier trained for MRI image is used for CT image~\cite{AAAI:2019:Chen:crossmodality}.

Besides radiology images, AI, in particular CV, can assist in diagnosis through several other means, such as electrocardiograms~\cite{IJCAI:2019:Hong:mina,AAAI:2019:Golany:pgans,IJCAI:2019:Zhou:atrial}, retinal imaging~\cite{AAAI:2019:Lim:ischemic,AAAI:2019:Zhao:glaucoma,AAAI:2019:Niu:retinal,KDD:2019:Tian:retinopathy,KDD:2018:Sugiura:glaucomatous}, and staining~\cite{AAAI:2019:Wu:g2c}.

In addition to using AI for disease diagnosis, there is also work which uses hierarchical probabilistic models to better understand physicians' diagnostic decision-making and reduce misdiagnoses in the clinical setting~\cite{IJCAI:2017:Guo:utterances}.

\paragraph{Clinical treatment}

Treatment recommendations for various diseases and conditions have also been made readily available by AI. 
At the most micro level is medication prescription. Shang et al. utilize graph neural networks (GNN) and Bidirectional Encoder Representations from Transformers (BERT), which is a state-of-the-art NLP pre-training model, to make medication recommendations~\cite{IJCAI:2019:Shang:medication}. Dual to medicine is the important problem of prescription surveillance. The misuse of antibiotics, for example, has become a severe problem worldwide. Beaudoin et al. design a prescription surveillance system that has been deployed at the Centre Hospitalier Universitaire de Sherbrooke~\cite{AIMAG:2014:Beaudoin:antimicrobial}.
Going one level up, treatment protocols often include multiple steps, for which reinforcement learning becomes a natural tool.
Guez et al. use fitted Q-iteration and randomized tree techniques to optimize deep-brain stimulation strategies for reducing seizure frequency and duration~\cite{IAAI:2008:Guez:epilepsy}.
There have been attempts to steer immune system adaptation using Monte Carlo Tree Search~\cite{IJCAI:2016:Kroer:immune}, and to control propofol anesthesia using Q-learning~\cite{IAAI:2010:Moore:propofol}.
Often in the treatment procedure, doctors order repeated clinical laboratory tests, but the timing and number of these tests are often open to personal judgement. This has also received some attention recently~\cite{AAAI:2015:Lasko:repeatedtests}.

The use of AI in clinical treatment also encompasses treatment preparation. IBM develops a tool to automatically generate a list of the patient's medical problems from historical EHR using NLP methods~\cite{IAAI:2015:Devarakonda:ehr}. New treatment protocols must be tested extensively before being used on patients. Arandjelovic uses statistical inference methods to adaptively reduce the time needed for the trials by sample size adjustment~\cite{AAAI:2015:Arandjelovic:clinicaltrial}, so that effective protocols can be used in practice sooner.

Also important in clinical treatment is counterfactual reasoning. When choosing between several treatment protocols based on historical data, we often want to know how each treatment would have worked on each individual. Since these data are typically unavailable, it is necessary to use causal inference to estimate the treatment effects. There is a rich literature on this topic~\cite{ICML:2017:Shalit:ite,ICML:2016:Johansson:cf,NeurIPS:2015:Swaminathan:cflearning}. For example, Hassanpour and Greiner propose an importance sampling re-weighing scheme with a representation learning model to estimate the individual treatment effects~\cite{IJCAI:2019:Hassanpour:cfregression}.
In the work of Atan et al., the proposed algorithm first learns an auto-encoder network to remove the bias in the original data, and then use another neural network to learn the optimal treatment policy~\cite{AAAI:2018:Atan:deeptreat}.

\paragraph{Clinical prediction}
Being able to predict clinical events can greatly assist the treatment of patients and improve operation efficiency in practice. Such prediction is made possible by the availability of EHR. The tasks in this direction can be roughly divided into diagnosis and disease progression prediction~\cite{AAAI:2015:Ghassemi:multivariate,IJCAI:2018:Qiao:pairwise} and patient subtyping. 
The problem is often formulated as sequential prediction on EHR data and the recently proposed approaches often involve some type of recurrent neural networks (RNN).

Diagnosis prediction might improve the quality of medical procedures~\cite{Medicine:2006:Chaudhry:preddiag}. This is also similar to the prediction of disease progression. The RETAIN model aims to build an interpretable RNN model which resembles the way human doctors using EHR to make diagnosis~\cite{NeurIPS:2016:Choi:retain}. Ma et al. use attention mechanisms to leverage the relationship between multiple visits and diagnosis prediction~\cite{KDD:2017:Ma:dipole}. Qiao et al. incorporate textual data such as clinical notes for diagnosis prediction using multimodal attentional neural networks~\cite{IJCAI:2019:Qiao:mnn}.
Esteban et al. study the clinical endpoint of kidney transplantation and use RNN to predict if the kidney will be rejected or the patient will die~\cite{ICHI:2016:Esteban:kidney}. 
Fu et al. use dictionary learning to make use of both labeled data and unlabeled data~\cite{IJCAI:2019:Fu:phenotyping}.
Choi et al. also predict the time duration to the next visit~\cite{MLHC:2016:Choi:doctorai}.

Patient subtyping attempts to group patients with similar disease progression together, so that a more personalized treatment plan can be carried out. Baytas et al. propose a time-aware long short-term memory (LSTM) network to address the heterogeneity of time intervals in patients' EHR data~\cite{KDD:2017:Baytas:subtyping}.

\paragraph{Hospital management}
There are a lot of operational challenges in hospital management that AI can help to address. For example, Rosemarin et al. propose an online deep learning-based scheduling algorithm for matching caregivers to patients in the emergency department~\cite{AAAI:2019:Rosemarin:ed}. Also in the emergency department setting, recent work used MIP to optimize the ambulance allocation~\cite{AAAI:2015:Saisubramanian:ambulance}.
As a routine practice, hospitals need to maintain an inventory of medical supplies and often need to share these supplies among them during large outbreaks. Researchers have used game-theoretic models to reason about the best sharing strategies~\cite{AAAI:2016:Lofgren:stockpiling}.

\paragraph{Transplant matching}
The compatibility of donors and recipients in organ transplant procedures is crucial to the survival of patients and success of the operation, and recent data-driven approaches have helped to assess this compatibility and make transplant matching recommendations. 
Yoon et al. propose a classification algorithm called ConfidentMatch~\cite{AAAI:2016:Yoon:organtransplant}, which predicts organ transplant success based on clinical and demographic traits of donor and recipient,
ConfidentMatch divides the feature space into clusters and train classifiers for each cluster. It was shown to surpass benchmark algorithms by several hundred patients in predicting transplant success.

When a patient needs an organ transplant, a donation from their relatives is sometimes the easiest option. However, the relatives often find themselves unable to do it due to incompatibility. This gives rise to the popularity of organ exchange platforms, where a patient and their incompatible donor can be matched with another patient-donor pair, if one patient's donor happens to be compatible with the other patient and vice versa. Although such matching problem has been studied in economics for a long time~\cite{QJE:2004:Roth:kidney,AER:2007:Roth:kidney}, national-level matching was practically infeasible until the work of Abraham et al.~\cite{EC:2007:Abraham:kidneyexchange}, whose algorithm was tested at the Alliance for Paired Donation, one of the leading kidney exchanges. This research spurred a series of works on the computational aspects of organ matching markets (e.g., ~\cite{IJCAI:2009:Awasthi:online,EC:2013:Dickerson:failure}). They have led to more efficient algorithms deployed on the kidney exchange platform at the United Network for Organ Sharing, which manages a US national organ transplant system that includes 69\% of the transplant centers in the country.

\paragraph{Drug development}
AI can be used to model the chemical interactions and properties relating to drug intake, which ultimately contributes to the development of new drugs that aid the general public. \cite{IJCAI:2018:Ma:drug} utilizes multi-view graph auto-encoders to learn similarity measures between different drugs. This is useful for inferring novel drug properties like side effects and interactions. In addition, both adverse reactions and adverse drug-drug interactions have been more explicitly studied~\cite{AAAI:2017:Xiao:adverse,AAAI:2017:Jin:drugdrug}, using latent Dirichlet allocation and multitask regression, respectively.
Drug repurposing, a popular technique, aims to find new functions of existing drugs, which is sometimes more cost-efficient than developing new drugs. Leveraging EHR and biomedical knowledge, some works discover drug families which were previously unknown for the target disease using knowledge graphs or solving LASSO problems ~\cite{AAAI:2019:Nordon:repurposing,KDD:2016:Kuang:repositioning}.

\paragraph{Mental health}
The importance of one's mental health on their physiology and general well-being has recently come to the forefront of medical focus. Various new AI applications bring similar techniques applied to other components of human health and physiology to mental afflictions and treatments, which were previously understudied. Chung et al. encode the fixation sequences captured from the visual scanning of patient's faces, and use LSTM on them to differentiate bipolar and unipolar patients~\cite{AAAI:2018:Chung:bipolar}. Brown et al. propose a sparse combined regression and classification formulation for Post-Traumatic Stress Disorder diagnosis based on peripheral physiology~\cite{AAAI:2015:Brown:ptsd}. More generally, there have been efforts to predict suicide risk based on medical history using a set of ordinal classifiers~\cite{KDD:2013:Tran:suicide} and to detect depression from social media posts using multimodal dictionary learning and transfer learning~\cite{IJCAI:2018:Shen:depression,IJCAI:2017:Shen:depression} and mobile phone usage~\cite{KDD:2017:Cao:deepmood}.

\paragraph{Well-being}
In today's fast-paced society, fatigue and other chronic symptoms are growing more and more common, even though these may not be clinically treatable symptoms. AI can also help in promoting the general well-being of the population. For example, a few works study and predict the sleep quality and sleepiness from daily logs and respiratory information using off-the-shelf classifiers as well as LSTMs~\cite{KDD:2019:Park:sleepquality,IJCAI:2019:Shinoda:sleepiness}. Aggarwal et al. use an unsupervised representation learning technique on daily physical activity and sleep patterns to predict health disorders~\cite{AAAI:2019:Aggarwal:activity}.

\subsubsection{Public health}
Public health is an important component of the healthcare system, which concerns itself with the disease prevention and health promotion of the whole community. Although it has a much wider definition, in this survey we focus on the epidemics monitoring, forecasting, and control, which have received the most attention from the AI community.

\paragraph{Epidemics monitoring} Dealing with epidemics is a top challenge in public health. The very first step is often the surveillance of epidemics, which lays the foundation for other tasks. Pei et al. propose an active surveillance method that actively selects areas to monitor in order to deal with the limited resources in practice~\cite{AAAI:2018:Pei:epidemic}.
Being able to detect disease outbreaks can often save enormous efforts and resources of the public health authority.
For example, Sadilek et al. mine Twitter data and use support vector machines (SVM) to detect events that may pose a public health hazard such as venues that lack permits or have contagious kitchen staff in an attempt to reduce foodborne illness~\cite{AAAI:2016:Sadilek:nemesis}. The system was deployed in Las Vegas. A follow-up work uses a similar supervised classifier on Google search history to identify potential users with foodborne illness and the source of the problem~\cite{Nature:2018:Sadilek:foodborne}.

\paragraph{Epidemics forecasting}
Assuming such data, the decision-maker might wish to predict the spread of the disease. A few works directly consider this as time series prediction, which uses the previous states of epidemics to predict the future~\cite{KDD:2019:Adhikari:epideep,PLOS:2015:Brooks:epidemics,IAAI:2019:Wang:defsi}. Other works use network epidemiology models to explicitly reason about the spread of disease across a social network, or an abstraction of a social network~\cite{KDD:2014:Beckman:isis,KDD:2018:Wang:metapopulation}. 

\paragraph{Epidemics control}
The epidemics control problem is frequently studied as the allocation of scarce resources, which reflects the unfortunate fact of the general lack of funding in the public health domain.
A few works study how to counter epidemics by intelligently choosing some network nodes to protect~\cite{AAAI:2018:Wilder:dynamic}. Some other works, such as that of Yadav et al. develop strategies to selectively educate homeless youth in Los Angeles to spread HIV-related information in order to reduce the risk of HIV infection~\cite{AAMAS:2017:Yadav:hiv}.
In another work, Killian et al. focus on the drug adherence problem in tuberculosis control in India~\cite{KDD:2019:Killian:tb}. Healthcare providers need to follow up with patients to make sure they adhere to the medication. Since the human resource is relatively scarce compared to the patient population, it is important to prioritize intervention to the ``riskiest'' patients, which Killian et al. attempt to do using an end-to-end decision-focused learning algorithm.

\subsubsection{Case study: raising awareness about HIV among homeless youth}
\paragraph{Target problem}
HIV brings a serious threat to public health. Homeless youth is a population particularly susceptible to HIV spread because of injection drugs and unsafe sexual activities. Therefore, raising awareness about HIV among homeless youth is a pressing problem. A few recent works~\cite{AAMAS:2017:Yadav:hiv,AAMAS:2016:Yadav:healer} used techniques from sequential decision making and influence maximization to raise the awareness of HIV prevention among homeless youth. This problem falls into ``community -- prescriptive'' under the AEC -- DPP framework.

\paragraph{Why is AI needed}
Community service providers routinely launch ``peer leader'' programs to teach selected youth about HIV prevention, hoping that these peer leaders will spread the information to other homeless youth. Since these programs are constrained by funding and human resources and cannot effectively reach out to every homeless youth, it is important to select the ``right'' set of peer leaders. In practice, service providers use degree centrality which essentially selects the most popular youth. However, this is not necessarily the best selection method. The AI intervention can directly optimize the information spread. 

\paragraph{Intervention overview}
The authors formulate a planning problem in terms of a partially observable MDP (POMDP), with the goal of influence maximization on a social network. Since the existing POMDP solvers do not scale to the size of the problem, Yadav et al. propose a hierarchical approach which decompose the POMDP into smaller ones. They solve these smaller POMDPs using Tree Aggregation for Sequential Planning which is based on a variant of the UCT algorithm (Upper Confidence Bounds applied to Trees) and then aggregate the results~\cite{AAMAS:2016:Yadav:healer}. At each time step, the planning algorithm selects a small set of homeless youth as the ``peer leader'' who will participate in the program at the service provider. 

\paragraph{Data used} 
The key parameters in this application are the social network connectivity, i.e. which homeless youths are friends with each other. This information was gathered in two ways. First, the authors developed a Facebook application which parses the youth's Facebook contact list to determine the friendship status. The authors also relied on the reports from the collaborating service partner who does interviews with the homeless youths.

\paragraph{Resources needed}
The authors collaborated with the homeless youth service providers, who are central to the success of the pilot study, as they recruited the youth and implemented the program for the peer leaders. Working with social work researchers also provides the necessary context and skills to communicate with the service providers and the youth.

\paragraph{Deployment status}
Starting in the spring of 2016, the authors launched three pilot studies in Los Angeles for a seven-month period, involving 173 homeless youths. The experiment compared two AI algorithms~\cite{AAMAS:2016:Yadav:healer,AAMAS:2017:Wilder:dosim} and the current practice. The result of the pilot studies showed that the AI interventions achieved 184\% more HIV prevention information spread than the current practice, and are also significantly better than the current practice in inducing behavioral changes in homeless youth. This suggests the great promise of future deployment.
That said, we were not able to find any public information about their continued deployment.

\paragraph{AI in retrospect}
The pilot study illustrates that the AI intervention effectively improves the spread of HIV awareness over the current practice. A prominent reason is that the AI algorithm can capture more global information in the network while the current practice is a greedy approach which leads to more redundant efforts. However, the study also showed that the AI algorithm has to be flexible enough to deal with the frequent incidents in human studies. If the AI algorithm outputs a certain youth to be the peer leader yet that youth was not able to show up to the program, then the algorithm needs to generate an alternative plan. The authors first address this problem by using heuristic solutions~\cite{AAMAS:2016:Yadav:healer}. This problem, along with the uncertainty of social network, is more systematically addressed in the follow-up work~\cite{AAMAS:2017:Wilder:dosim}.

\subsubsection{Other works}

The reader is referred to the survey by Yu et al. for an overview of the AI for clinical health outside the major AI literature~\cite{Nature:2018:Yu:healthsurvey}.
\subsection{AI for Combating Information Manipulation}
\label{sec:info}
\begin{table}[ht]
    \centering
    \begin{tabular}{|c|c|C{10cm}|}
    \hline
    \multirow{3}{*}{Agent} & Descriptive & Personalized sentiment analysis (Emotion inference from social media \cite{AAAI:2014:Yang:emotions})  \\
    \cline{2-3}
     &  Predictive & Link prediction with personalized social influence \cite{KDD:2019:Xu:link} \\
    \cline{2-3}
     &  Prescriptive & Trustworthiness evaluation \cite{IJCAI:13:Fang:advisors} \\
    \hline
    \multirow{3}{*}{Environment} & Descriptive & Political ideology detection \cite{IJCAI:2017:Chen:political} \\
    \cline{2-3}
     &  Predictive & Misinformation prediction \cite{IJCAI:2017:Yu:misinformation} \\
    \cline{2-3}
     &  Prescriptive & Improving robustness of trust systems \cite{AAMAS:2015:Wang:robustness} \\
    \hline
    \multirow{3}{*}{Community} & Descriptive & Modeling planned protest \cite{IAAI:2015:Muthiah:protest} \\
    \cline{2-3}
     &  Predictive & Blame detection \cite{AAAI:2019:Liang:blame} \\
    \cline{2-3}
     &  Prescriptive & Coverage Centrality Maximization in Undirected Networks \cite{AAAI:2019:DAngelo:coverage}\\
    \hline
    \end{tabular}
    \caption{AEC and DPP categorization of research on AI for combating information manipulation.}
    \label{tab:information}
\end{table}

Combating information manipulation has become a prominent issue in recent years with the rise of new media channels. AI can play a significant role in ensuring that citizens around the world are utilizing trustworthy information and preventing systematic dispersion of false news. We believe there is much work to be done in this domain. We show several example works according to the AEC and DPP structure in Table~\ref{tab:information}. In this subsection, we discuss the works on two major challenges to information validation: misinformation and fraud.

\subsubsection{Combating misinformation}

Much of AI for combating information manipulation aims at spreading awareness of misleading information. This has become an increasingly prominent issue in browsing the web in recent years, with the rise of fake news and the creation of potentially malicious technologies like Deepfakes. 
In the sequel, we first discuss the misinformation detection which has been the center of attention in the literature. We also introduce the research on clickbait detection as a specialized topic and the detection of social bots which are an important means of fake news campaigns.

\paragraph{Misinformation detection} 
As a preventative measure against the spread of false information on the Internet, the first step is to detect its presence on pages like news channels and social media platforms.
Since the seminal work of Castillo et al. on the information credibility on Twitter~\cite{WWW:2011:Castillo:twitter}, social media rumor detection has received wide attention from the literature~\cite{KDD:2012:Yang:rumor,ICDM:2013:Kwon:prominent,WWW:2015:Zhao:enquiry}, leveraging classifiers like decision tree, SVM, and random forest.
Recent works~\cite{IJCAI:2016:Ma:rumorrnn,IJCAI:2017:Yu:misinformation,IJCAI:2018:Qian:fakenews} use RNNs and convolutional neural networks (CNN) to automatically extract key features from social media posts to identify misinformation, and have achieved practical early detection of misinformation in experimental results on large datasets from Twitter and Weibo. 
These works typically rely on the dynamics of the social media posts, such as the post's popularity, the source's credibility, and the timing of retweets, to perform classification. 

In the natural language processing community, lots of works focus on analyzing the text of the fake news itself~\cite{COLING:2018:Perez:fakenews}, with well-received public benchmark datasets~\cite{ACL:2017:Wang:fakenewsdata,ACL:2014:Vlachos:fakenewsdataset}. The CSI model proposed by Ruchansky et al., which consists of three modules Capture, Score, and Integrate, combines network features and textual features to detect fake news~\cite{CIKM:2017:Ruchansky:csi}. In a recent work, Zellers et al. present a system which can generate fake news which appears more credible than human-written news using deep neural network-based language models~\cite{NeurIPS:2019:Zellers:grover}. Their system can also detect fake news with higher accuracy than other existing classifiers.

\paragraph{Clickbait headlines}
Another type of misinformation is clickbait headlines, which attract the reader's attention with exaggerated or false information. There have been a few attempts to detect clickbait headlines~\cite{ASONAM:2016:Chakraborty:clickbait,AAAI:2016:Biyani:8secrets,IJCAI:2017:Wei:ambiguous}. Yoon et al. use a deep hierarchical encoder to detect incongruity between the news headline and the body text~\cite{AAAI:2019:Yoon:incongruity}. Liu et al. use a neural matching model to explain ambiguous headlines of online news~\cite{IJCAI:2018:Liu:ambiguous}.

\paragraph{Bot detection}
Of particular significance in the disinformation campaigns is social bots. These are social media accounts managed by algorithms that automatically post or re-post information. In the hands of disinformation campaigners, these bots are a powerful tool to spread and reinforce the distorted information. Detecting such bots has been a popular topic in AI research recently~\cite{TKDD:2014:Yang:sybil,AAAI:2014:Hu:spammer}. As we will study in Section~\ref{sec:infocase}, Varol et al. developed the first public bot detection tool based on classification algorithms with carefully engineered features~\cite{WSM:2017:Varol:botometer,CACM:2016:Ferrara:bots}.
Different from these supervised approaches, Chavoshi et al. use a warped correlation finder to detect bot accounts that have correlated activities~\cite{ICDM:2016:Chavoshi:debot}.

\subsubsection{Combating fraud}
In addition to fake news, AI has also been applied to detect misleading or dishonest information, also known as fraud detection. We review two most notable forms of fraud: reputation fraud and financial fraud.

\paragraph{Reputation fraud}
Reputation fraud aims to distort the public opinion on some targets such as goods and services. It usually takes the form of posting online malicious ratings, reviews, and knowledgebase entries. 
It is also related to the shilling attacks in recommender systems~\cite{KDD:2012:Wu:shilling}, and trustworthiness evaluation~\cite{IJCAI:2013:Fang:advisors}.
The initial attempt by Jindal and Liu~\cite{WSDM:2008:Jindal:opinionspam}, who used logistic regression to detect spam reviews, spurred a flurry of works on reputation fraud detection.
The first line of work relies on the textual review to classify malicious reviews. For example, Ott et al. use naive Bayes and SVM to find deceptive opinion spams~\cite{ACL:2011:Ott:imagination}. Another approach classifies spam information based on the user's historical behavior, e.g. previous reviews, trajectories, and average ratings~\cite{WSDM:2008:Jindal:opinionspam,ASONAM:2014:Lin:sequence,KDD:2013:Mukherjee:footprint}, using similar off-the-shelf classifiers as well as techniques like learning to rank~\cite{NOW:2009:Liu:learningtorank}. Some works combine the two approaches~\cite{WWW:2016:Kumar:wiki}.
Mukherjee et al. investigate the spam review detector on Yelp with the aforementioned learning algorithms, and discover that the implementation is closer to the latter approach~\cite{WSM:2013:Mukherjee:yelp}. Finally, a third approach is to detect fake reviews based on the interactions among reviewers as well as the review objects~\cite{KDD:2016:Hooi:fraudar,WSM:2013:Fei:burst} using techniques from topics such as graph mining and probabilistic graphical models.
For example, Xu et al. use online learning to detect reputation fraud campaigns and characterize the interactions between the campaigned spammers~\cite{IJCAI:2017:Xu:fraud}.
Rayana and Leman combine the three approaches together in a unified framework to detect suspicious users and reviews using inference methods on graphical models~\cite{KDD:2015:Rayana:collective}.

\paragraph{Financial fraud}
Financial fraud is the dishonest and covert attempt to escape the oversight of law or other regulations in the pursuit of monetary gains.
There have been existing works to fight against business procurement fraud using a number of techniques from NLP, social network analysis and online learning~\cite{IAAI:2015:Dhurandhar:ibm}, real money trading in online games using multi-view attention networks~\cite{KDD:2019:Tao:netease}, and cash-out detection in financial services using neural networks with hierarchical attention mechanism ~\cite{AAAI:2019:Hu:cashout}. Some of these have even been deployed~\cite{IAAI:2015:Dhurandhar:ibm,KDD:2019:Tao:netease}.
In healthcare, fraud in insurance claims has been a central topic for years. Chandola et al. use methods from topic modeling, social network analysis, and statistical process control to classify fraudulent claims~\cite{KDD:2013:Chandola:claim}. A graph-based approach for detecting healthcare fraud has been put into use~\cite{IAAI:2015:Liu:graph}. Recent work exploits the temporal nature of fraud identification, which, combined with human-engineered features, yields a high rate of fraud detection~\cite{IAAI:2019:Qazi:Tpattern}.

\subsubsection{Case study: detecting social bots on Twitter}
\label{sec:infocase}
\paragraph{Target problem}
Social networking platforms have become a main news and information source for many users. Some users, however, abuse these platforms to spread misinformation in accordance with some potentially malicious agenda.
A subset of such users create automated or ``bot'' accounts that automatically post content and form false social relationships.
This produces an exigence for automated detection of social media bots, which is crucial to helping the general public make more informed decisions about which information sources to trust.
One such tool is Botometer, which was introduced and developed in a number of works~\cite{WWW:2016:Davis:botornot,WSM:2017:Varol:botometer,AAAI:2020:Yang:socialbot}. This problem falls into ``community -- descriptive'' under the AEC -- DPP framework.
    
\paragraph{Why is AI needed}
Prior to the existence of the authors' solution, Botometer, the standard approach to detection of malicious bots on social media platforms was manual validation from either the platform's users or administrators. 
    Botometer reduces the amount of manual inspection required of human users in analyzing Twitter posts, and significantly out-scales the manual approach in terms of accuracy. 
    
\paragraph{Intervention overview}
    Botometer, first gathers data from the social media site Twitter via its public API. It then extracts over one thousand features relating to the user in question's account, friends, network, content, and sentiment. Botometer finally employs a random forest classifier model trained on a public dataset (a subset of which is manually annotated as human or bot) with thousands of verified bot accounts, their posted tweets, and tweets mentioning them. The classifier outputs a probability value representing how likely the given account is a bot.

    Public usage of the deployed application involves inputting the screen name of a specific Twitter user, and the system subsequently outputs a bot likelihood score for that account. This is all done by interacting with their web application at \url{https://botometer.iuni.iu.edu}.
    
\paragraph{Data used} 
    Gathering of Twitter data is the lifeblood of this project. 
    As such, Botometer collects tweets on 14 million user accounts satisfying a certain activity threshold, relying on the Twitter Search API. 
    Twitter user datasets give the authors current predictions about the percentage of active Twitter accounts which are bots, as well as the interactions between bots with other accounts (whether human or bot).
    
\paragraph{Resources needed}
    During the Botometer classifier's training process, the authors utilized an annotated dataset created by four volunteers who manually assessed the bot status of each user in a subset of the original dataset of users.
    Other than this, the project's implementation and maintenance are fairly self-contained, mainly relying upon the Twitter API.
    
\paragraph{Deployment status}
    Since its release in 2014, Botometer has been open to public use and is now available at the aforementioned site. 
    The authors report that the tool serves hundreds of thousand requests daily~\cite{Nature:2018:Shao:socialbot}.
    Due to its automated nature and method of usage, the current tool does not require significant maintenance of the authors. However, the authors are still developing and improving upon the Botometer classifier, and users of the application can continuously provide feedback for the system regarding the accuracy of specific classifications. The tool is also used in other projects such as Hoaxy\footnote{\url{https://hoaxy.iuni.iu.edu/}}, a visualization and fact-checking website.
    
\paragraph{AI in retrospect}
    The AI intervention in general was quite accurate, achieving strong consistency with the manually validated dataset. However, some aspects of Botometer's classification were not so successful. For instance, some Twitter accounts are run by both human and automated software. In these boundary cases exhibiting behaviors of both human and bot activity, it was more difficult for Botometer to accurately and confidently predict whether the account is a bot or not. 
    Another area of weakness of the AI intervention was its susceptibility to false positive classifications. For instance, Botometer tended to classify human accounts who post frequently from externally connected applications, or in multiple languages as bots. To tackle the latter case, the authors modified the classifier to ignore language-dependent features.

\subsubsection{Other works}

The reader may refer to the work of West and Bhattacharya for a survey of AI in fraud detection~\cite{CS:2016:West:fraudsurvey}.
\subsection{AI for Social Care and Urban Planning}
\label{sec:socialcare}

\begin{table}[h]
    \centering
    \begin{tabular}{|c|c|C{10cm}|}
    \hline
    \multirow{3}{*}{Agent} & Descriptive & Informing citizens (Increasing financial literacy among Bangalore drivers \cite{ICTD:2017:Mehra:prayana})  \\
    \cline{2-3}
     &  Predictive &  \\
    \cline{2-3}
     &  Prescriptive & Public service allocation to individuals (SMS public forum platform \cite{KDD:2011:Melville:voice}) \\
    \hline
    \multirow{3}{*}{Environment} & Descriptive & Poverty mapping \cite{AAAI:2015:Xie:mapping}  \\
    \cline{2-3}
     &  Predictive & Disaster forecasting (Severe hail prediction \cite{IAAI:2015:Gagne:hail}) \\
    \cline{2-3}
     &  Prescriptive & Public infrastructure improvement (Optimizing water pipe maintenance \cite{IJCAI:2013:Yan:waterpipe})\\
    \hline
    \multirow{3}{*}{Community} & Descriptive & Urban statistic collection (Urban resident recognition to study migration \cite{AAAI:2018:Wang:recognition}) \\
    \cline{2-3}
     &  Predictive & Inference of social ties (Learning social structure of hidden populations \cite{AAAI:2016:Chen:socialties}) \\
    \cline{2-3}
     &  Prescriptive & Public service optimization (Improving homelessness services allocation \cite{AAAI:2019:Kube:homelessness}) \\
    \hline
    \end{tabular}
    \caption{AEC and DPP categorization of research on AI for social care and urban planning.}
    \label{tab:socialcare}
\end{table}

Social care and urban planning provide the services and infrastructure that citizens around the globe rely on in a functioning and supportive community. Unfortunately, these public sectors are sometimes restricted in their resources, rendering their residents unprepared for daily life and work. AI is a promising tool to intelligently save or allocate the scarce program resources, e.g. helping put an end to such humanitarian crises as the Flint water crisis, or providing more wide-reaching rehabilitation services to the homeless. Table~\ref{tab:socialcare} shows several sample works on AI for social care and urban planning, categorized using the AEC and DPP structure. In this subsection, we outline two major areas -- public services and crisis management -- which have been improved with the application of AI.

\subsubsection{Public services}

A primary goal of AI for social care and urban planning is the provision of adequate public services for members of a community. 
Most of the existing works in the AI community can be grouped into the following three categories: collecting urban data for urban planning, optimizing public infrastructure in major metropolitan areas, as well as establishing more basic services to underserved communities.

\paragraph{Urban data collection} Urban planners must gather various statistics on the community they are serving, to better infer how to improve the lives of community members. The wealth of data affords city planners with a wide variety of possibilities. For example, check-in locations on social media and online map search queries are useful for identifying points of interest and planning urban planning and investments~\cite{KDD:2018:Zhou:wechat,KDD:2018:Sun:regionofinterest}.
Data are also helping cities pay more attention to the neighborhoods that need it the most~\cite{KDD:2015:Green:revitalization}.
Recent work in urban data collection aims at gathering more unbiased data, and extracting more useful information from smaller amounts of data (e.g., avoiding usage of tools like CCTV in favor of satellite imagery). 
Xie et al., for instance, utilize transfer learning to extract features from satellite imagery which can predict poverty, aggregating poverty data in developing countries~\cite{AAAI:2015:Xie:mapping}. Transfer learning is also useful when the target city's service or infrastructure is just built but abundant data are available at similar cities~\cite{KDD:2016:Wei:transfer}.
\cite{AAAI:2018:Wang:recognition} attempts to model migration patterns via a cross-domain deep learning model. 

\paragraph{Infrastructure optimization} The design and maintenance of urban infrastructure are a main functionality of city governance. In the AI literature, partly due to the availability of systematic records and data, there have been some works concerning water consumption and provision~\cite{KDD:2013:Kermany:meter,KDD:2018:Kumar:watermain}. Yan et al. aim to optimize water pipe replacement and rehabilitation by examining more novel approaches using binary classifiers and self-exciting stochastic processes for predicting water pipe failure~\cite{IJCAI:2013:Yan:waterpipe}. The Flint water crisis is a well-known problem in the US. Some attempt to use data science to understand the water contamination in Flint~\cite{KDD:2017:Chojnacki:flint}. In the case study of this section, we will detail the subsequent effort of replacing the water pipes in Flint~\cite{KDD:2018:Abernethy:flint}.

\paragraph{Social work} On top of infrastructure, public workers are also interested in the optimal allocation of public resources and the provision of such services to underserved populations. Predictive modeling is a major technique used by the literature along this line. Existing works span from using ML models such as gradient boosting to help vulnerable renters~\cite{COMPASS:2019:Ye:tenants}, to combining learning with integer programming for peer-to-peer rental recommendation ~\cite{KDD:2017:Fu:p2prental} to modeling population migration with XGBoost and neural networks~\cite{COMPASS:2018:Robinson:migration} to using random forests to predict a poverty heat map which guides cash transfer campaigns for poverty relief~\cite{KDD:2014:Abelson:cashtransfer}. In a recent work of Kube et al., the authors attempt to choose the optimal allocation of various homelessness services using counterfactual inference, and specifically Bayesian additive regression trees~\cite{AAAI:2019:Kube:homelessness}.

A specific topic that has received much attention is food insecurity, which has its presence in the least developed regions as well as the wealthiest countries.
Optimizing the efficiency and fairness of food bank and food rescue operations is a step towards mitigating this issue.
Shi et al. use predictive modeling such as the stacking model to predict the success or failure of a food rescue, and use data-driven optimization to improve the operation of a food rescue organization in Pittsburgh, US~\cite{IAAI:2020:Shi:foodrescue}. Aleksandrov et al., in collaboration with a food bank in Australia, propose two online fair division mechanisms which address how food bank might allocate the received food to charities~\cite{IJCAI:2015:Aleksandrov:foodbank}. Silvis et al. developed a mobile app that uses reinforcement learning to select users to invite to events that have leftover food~\cite{KDD:2018:Silvis:pittgrub}.

\subsubsection{Crisis management}

In addition to public services, citizens in urban environments also rely on their community's general preparedness for crises, such as natural disasters. We introduce the research efforts in two major directions: disaster forecasting and disaster response.

\paragraph{Disaster forecasting} The first step toward preparedness is predicting when natural disasters may strike. For example, Gagne et al.~\cite{IAAI:2015:Gagne:hail} use standard machine learning models like gradient boosting trees and linear regression to predict upcoming hail storms and their sizes. 
Chen et al~\cite{IJCAI:2013:Chen:forecast} focus on the problem of hurricane prediction. Radke et al.~\cite{IJCAI:2019:Radke:firecast} use CNN to predict the spread of wildfire.
Disaster forecasting is typically studied as extreme or rare event prediction in the literature~\cite{KDD:2019:Ding:extreme, IJCAI:2013:Chen:forecast}. In this front, Ding et al.~\cite{KDD:2019:Ding:extreme} improve the prediction performance of deep neural networks by using a novel loss function called the extreme value loss. 
\cite{KDD:2014:Avvenuti:EARS} makes use of NLP techniques to mine Twitter feeds real-time as a source of sensor data, to predict the occurrence of earthquakes.

\paragraph{Disaster response} Preparing measures such as search, rescue, and evacuation in response to the inevitable occurrence of disasters is equally important as being prepared for when they strike.
With the forecasting mentioned above, governments can plan for evacuation by solving a graph flow problem~\cite{AAAI:2016:Romanski:benders}. This is often coupled with the preparation in transportation infrastructure network about which link to enhance, solved as a stochastic optimization problem~\cite{AAAI:2015:Schichl:predisaster,AAAI:2016:Wu:transportation} or using Benders decomposition~\cite{AAAI:2016:Kumar:evacuation}. This is also related to how to dispatch emergency response vehicles, as studied by Ghosh and Varakantham using a mathematical programming approach~\cite{AAAI:2018:Ghosh:erv}. 

Such planning often relies on the information about the emergency situation and human mobility following the disaster. 
As the very first step, the decision-maker requires immediate knowledge of the mere occurrence of the disaster. Avvenuti et al. use NLP techniques to mine Twitter feeds in real-time as a source of sensor data, to detect the occurrence of earthquakes~\cite{KDD:2014:Avvenuti:EARS}.
Second, another important piece of information is damage estimation. Rudner et al. propose a novel CNN-based approach to use satellite images to detect and segment affected buildings after flooding~\cite{AAAI:2019:Rudner:multi3net}.
Third, human mobility data is crucial to public management.
Leveraging data from the global positioning system, Song et al. simulate the human mobility pattern following large disasters such as earthquakes in Japan~\cite{AAAI:2014:Song:simulator,AAAI:2015:Song:simulator}. Parikh et al. model the human behavior after a nuclear detonation~\cite{AAMAS:2013:Parikh:nuclear}.

Recently, the use of UAVs has become more commonplace in a number of social service sectors. In disaster response, UAVs can be programmed to collect information about an unknown environment~\cite{IAAI:2012:DelleFave:maxsum}, and collaborate with and crowdsource from humans to guide their actions~\cite{AAMAS:2015:Ramchurn:hac}.
Advancing from the full information scenario, Wu et al.~\cite{IJCAI:2016:Wu:coordinating} model the problem of coordinating human-UAV teams in disaster response as a POMDP, which integrates information gathering with task planning to guide both humans and UAVs in disaster response.

\subsubsection{Case study: the search for lead pipes in Flint, Michigan}

\paragraph{Target problem}
The water crisis in Flint, Michigan is a widely known man-made urban disaster that has hit the headlines since 2015. Throughout the city of Flint, the lead water pipes suffered from corrosion, which led to the contamination of drinking water and serious public health issue. The city decided to replace the affected lead water pipes. 
A group of researchers at the University of Michigan used predictive modeling and active learning methods to guide the replacement procedure~\cite{KDD:2017:Chojnacki:flint,KDD:2018:Abernethy:flint}. This problem falls into ``community -- prescriptive'' under the AEC -- DPP framework.

\paragraph{Why is AI needed}
The default option is to inspect each household's water pipes and replace them if needed. However, such inspection, by digging into the ground, requires too much funding if carried out on all housings in the city. Meanwhile, it is also desired that the housings with unaffected water pipes never get inspected, because they do not require replacement and thus the inspection cost is wasted. 
    Therefore, it is important to identify the ``high-risk'' housings in advance so that the limited budget can be applied to the housings that need it the most. This is the goal of the authors' AI intervention.
    
\paragraph{Intervention overview}
    The authors first build a predictive model which, given certain features of the household, predicts the likelihood that this household has contaminated water pipes. This predictive model consists of an XGBoost classifier and a hierarchical Bayesian model. The latter is a relatively novel component which resolves the dilemma that each region in Flint has different amount of information. The Bayesian model allows the authors to leverage the data from all regions and makes the result of regions with less information closer to the city average.   
    
    Based on this predictive model, the authors develop an active learning framework to guide which household should be inspected in a sequential fashion. The algorithm is a variant of the importance weighted active learning, which adaptively samples data from regions with high uncertainty and incorporates the acquired data into the predictive model.

\paragraph{Data used} 
In this project, the quantity and quality of data are essential to the result of the AI intervention. Both of them posed a significant challenge and the authors took initiative to address this challenge. The predictive model needs the water pipe status of as many households as possible, yet at the beginning the authors only have 250 data points from the city of Flint. One reason for such scant data is that other data sources proved to be unreliable. Even the 250 available data points are not sampled (nearly) randomly. The authors worked with the city to use a cheaper method of collecting data, which entails working with other contractors in a way that the authors can have autonomy in which data points to collect.

\paragraph{Resources needed}
The authors worked closely with the city government, the various contractors who do the actual work of replacing pipes, and researchers in related domains. The authors also made technical efforts to facilitate the collaboration. For example, they developed a mobile and web app to help the contractors log the data.

\paragraph{Deployment status}
The proposed AI-enhanced replacement plan was adopted by the city of Flint, and was successful. As reported by Madrigal, in 2007 the city of Flint inspected 8833 homes and replaced the pipes of 6228 of them, which means the algorithm achieved an impressive 70\% accuracy in practice~\cite{Atlantic:2019:Madrigal:flint}.
There have been some setbacks after the AI intervention's initial success in 2016 - 2017. However, the setbacks were largely due to factors beyond the researchers' control. Fortunately, the AI intervention is back to running in 2019. The reader can refer to the piece by Madrigal~\cite{Atlantic:2019:Madrigal:flint} for more detailed information.

\paragraph{AI in retrospect}
The accuracy of the predictive model validated in the deployment during 2016 - 2017. In addition, the low performance of an alternative brute force method that the city of Flint used in 2018 further corroborated the efficacy of the AI intervention. However, one reason of the setback in 2018 is that the selective inspection caused confusion in the neighborhood: ``you did my neighbor's pipe but didn't do mine, why is that?'' The AI intervention could have considered making the algorithm more explainable to address the issue regarding the perception of fairness.
\subsection{AI for Public Safety}
\begin{table}[ht]
    \centering
    \begin{tabular}{|c|c|C{10cm}|}
    \hline
    \multirow{3}{*}{Agent} & Descriptive &   \\
    \cline{2-3}
     &  Predictive & Identifying risk for police officers \cite{KDD:2016:Carton:police} \\
    \cline{2-3}
     &  Prescriptive &  \\
    \hline
    \multirow{3}{*}{Environment} & Descriptive &   \\
    \cline{2-3}
     &  Predictive & Crime network prediction (Predicting offender networks \cite{KDD:2014:Tayebi:cooffence}) \\
    \cline{2-3}
     &  Prescriptive & Improving law enforcement deployment \cite{IJCAI:2019:Chase:lawenforcement} \\
    \hline
    \multirow{3}{*}{Community} & Descriptive & Investigation datasets (Hotel recognition dataset for trafficking investigations \cite{AAAI:2019:Stylianou:hotel}) \\
    \cline{2-3}
     &  Predictive & Crime detection (Detecting pickpocketing suspects \cite{KDD:2016:Du:pickpocket}) \\
    \cline{2-3}
     &  Prescriptive & Violence reduction \cite{KDD:2014:Shakarian:gang}\\
    \hline
    \end{tabular}
    \caption{AEC and DPP categorization of research on AI for public safety.}
    \label{tab:publicsafety}
\end{table}

With criminals becoming increasingly sophisticated and crimes still prevalent in many regions around the world~\cite{UNODC:UN:crime}, public safety poses a major challenge for the modern society.
Thus, there is an urgent need to improve the efficacy and trustworthiness of law enforcement systems, for which AI could be part of the solution.
As shown in Table~\ref{tab:publicsafety}, the research efforts on AI for public safety have mostly focused on the environment and community categories. 
According to whether the AI intervention is facing the public, criminals, or itself, we cluster the existing works into three topics: crime and event detection, combating crimes, and improving law enforcement operations.

\subsubsection{Crime and event detection}

Public surveillance as a law enforcement approach is embroiled in controversy. We share a lot of concerns about it, but as we mentioned in the introduction, we focus on its use in good hands. 
The person re-identification problem, that is, matching the images of the same person in different settings has received much attention from the vision community~\cite{IJCAI:2016:Wang:reid,AAAI:2018:Li:reid,AAAI:2018:Huang:reid}. Albanese et al. consider finding unexplained events in the video~\cite{IJCAI:2011:Albanese:unexplained}. Crowd prediction is closely related to public surveillance and has important implications to transportation and urban planning in addition to public safety.
Most works use CNN-based architecture to capture the spatial relationship. As for the temporal dependency, some directly use a mixed architecture of CNN and RNN~\cite{IJCAI:2018:Zonoozi:crn}, while others attempt to encode temporal information in the input to the pure convolutional network~\cite{AAAI:2017:Zhang:stresnet,AAAI:2019:Lin:deepstn+} because a mixed architecture is typically difficult to train.

Surveillance is not only about video cameras and images. The public and social media, for example, provide a rich source of information. Khandpur et al. use news and Twitter data to predict airport threats~\cite{IAAI:2017:Khandpur:airport}. The EMBERS system, a similar system to~\cite{IAAI:2017:Khandpur:airport}, predicts civil unrest events from news and social media data~\cite{KDD:2014:Ramakrishnan:embers}. 
EMBERS has been deployed in 10 countries since 2012.
Finally, in today's society everyone inevitably leaves their digital traces in many places.
Du et al.~\cite{KDD:2016:Du:pickpocket} detect pickpocketing suspects by extracting features from large-scale public transit records using both supervised and unsupervised techniques.

\subsubsection{Combating crimes}

The natural next step in safety enforcement is to prevent the occurrence of crimes after their detection and to reduce the severity of their impact on innocent citizens. One recent work constructs a dataset of 1 million images of 50000 hotels worldwide, in an effort to identify the location of the victims of human trafficking~\cite{AAAI:2019:Stylianou:hotel}. In the sequel, we elaborate on two directions that the literature has focused on: the strategic interaction with criminals and the study of the criminal network.

\paragraph{Strategic interactions}
Zhang et al.~\cite{AAAI:2019:Zhang:interdiction} attempt to reduce the frequency of violent crimes in urban environments with their proposed defender-adversary model, NEST, and an incremental strategy generation algorithm. The approach's effectiveness is experimentally shown for real networks up to the size of Manhattan.
The defender-adversary model used in this work, known as the security game, has been studied extensively in a number of works. Existing works include strategically countering attacks on ferries~\cite{AAMAS:2013:Fang:ferry}, cargo ships~\cite{AAMAS:2012:Jakob:piracy,AAAI:2018:Wang:captainjack}, ports~\cite{AAMAS:2012:Shieh:protect}, public infrastructure~\cite{IJCAI:2017:McCarthy:threat,AAMAS:2008:Pita:laairport}, border control~\cite{AAAI:2016:Guo:network}, traffic and transportation~\cite{IJCAI:2017:Rosenfeld:traffic,IAAI:2012:Yin:trusts,IAAI:2013:Varakantham:rtn,IAAI:2014:Brown:streets}.

\paragraph{Criminal network}  Gang violence poses a direct threat to public safety in a number of countries, but it exhibits a strong criminal network not commonly seen in other types of crimes.  Tayebi et al. utilize supervised learning techniques to extract social and geographical features to classify and predict potential networks of offenders who have committed crimes together~\cite{KDD:2014:Tayebi:cooffence}.
Also focusing on co-offender networks, Shaabani et al. use the network information to identify criminals~\cite{KDD:2015:Shaabani:gang}. 
Based on such networks, Shakarian et al. leverage techniques from influence maximization to encourage gang members to leave their gang~\cite{KDD:2014:Shakarian:gang}.

\subsubsection{Improving law enforcement operations} 
Optimizing law enforcement makes police operations as a whole more efficient and capable in keeping communities safe. 
Carton et al.~\cite{KDD:2016:Carton:police} attempt to augment the existing Early Intervention System (EIS), which is used for identifying police officers who are at risk of adverse events. The work could potentially reduce not only police casualties in events like shootings, but also police misconduct against civilians, like racial profiling or incidents of brutality.
The work of Chase et al.~\cite{IJCAI:2019:Chase:lawenforcement} aims to minimize response times of law enforcement by optimizing response time failure risk as a MILP and using Sample Average Approximation and Iterated Local Search. 

\subsubsection{Case study: identifying police officers at risk of adverse events}

\paragraph{Target problem}
Police adverse incidents such as the misuse of force pose an enormous challenge to the police reputation and effective operations in today's political climate.
A group of researchers used predictive modeling to develop an early intervention system that can predict such adverse incidents so that proper intervention can be carried out to prevent them from happening~\cite{KDD:2016:Carton:police}. This problem falls into ``agent -- predictive'' under the AEC -- DPP framework.

\paragraph{Why is AI needed}
Recognizing the severity of the problem, most police departments in the US have an EIS in place. However, these EISs are typically rule-based with rules designed by some experts, experienced yet admittedly arbitrary. It is not easy for a particular police department to adjust the rules according to its specific situation. The simple rule-based systems are also prone to gaming. A machine learning-based EIS could alleviate these shortcomings by making the model more accurate, customizable, and more robust to gaming. However, we would like to argue that whether the AI-based EIS can deliver these benefits is subject to doubt. Nevertheless, it is worthwhile to study this project to see how AI can function in this setting.

\paragraph{Intervention overview}
The authors formulated the problem as a binary classification task: whether a police officer will have an adverse event within one year from now. Most of the technical effort was focused on feature generation. A total of 432 features include information on the historical frequencies of adverse incidents, the training history, and the patrol neighborhood of a given officer. Several time series, normalized, or high-order features based on the aforementioned features are also included. The authors used several off-the-shelf classifiers such as AdaBoost, random forests, logistic regression, and SVM, and found that random forests performed the best. The authors also performed an exploratory predictive analysis about which dispatch event might lead to adverse events, and a similar random forest approach was used.

\paragraph{Data used} 
The authors obtained data on adverse incidents from the collaborating police departments. Also obtained from the police department include other data such as dispatch events, complaints, arrests, employee records, etc. They are used as input features in the predictive model. 

\paragraph{Resources needed}
This project was conceived under the White House's Police Data Initiative. Having a high-level endorsement probably ensured the resources for the collaboration with not just the partnering police department in~\cite{KDD:2016:Carton:police} but also several police departments across the country.

\paragraph{Deployment status}
The researchers on the project reported that such data-driven EIS was implemented or is currently being used at the police departments in multiple cities including Charlotte-Mecklenburg, Nashville, San Francisco and so on. The authors have also licensed the work to a commercial company to scale up deployment.\footnote{Benchmark Analytics. \url{https://www.benchmarkanalytics.com/}}

\subsection{AI for Transportation}
\begin{table}[h]
    \centering
    \begin{tabular}{|c|c|C{10cm}|}
    \hline
    \multirow{3}{*}{Agent} & Descriptive & Real-time parking availability \cite{KDD:2018:Rong:realtime}  \\
    \cline{2-3}
     &  Predictive & Predict arrival time~\cite{AAAI:2018:Wang:arrive} \\
    \cline{2-3}
     &  Prescriptive & Transportation recommendation~\cite{AAAI:2019:Liu:multimodal} \\
    \hline
    \multirow{3}{*}{Environment} & Descriptive & Traffic detection~\cite{IJCAI:2013:Joshi:acoustic}  \\
    \cline{2-3}
     &  Predictive & Traffic forecast~\cite{IJCAI:2018:Yu2018:stgcn}\\
    \cline{2-3}
     &  Prescriptive & Bike-share rebalancing~\cite{AAAI:2015:Omahony:citibike}\\
    \hline
    \multirow{3}{*}{Community} & Descriptive & Multi-agent transportation representation (Testbed for air traffic controllers \cite{AAAI:2010:Schurr:air}) \\
    \cline{2-3}
     &  Predictive & Bike-share demand prediction \cite{} \\
    \cline{2-3}
     &  Prescriptive & Route optimization~\cite{AAAI:2016:Cao:arrival} \\
    \hline
    \end{tabular}
    \caption{AEC and DPP categorization of research on AI for transportation.}
    \label{tab:transportation}
\end{table}

The number of vehicles operating on public roads increases each day, and with it, the amount of congestion and vulnerability to fatal accidents or injury. Many developments in transportation have served to improve efficiency of existing transportation methods, thereby mitigating the aforementioned issues. AI can be utilized to help accomplish this, which has already been done to a great extent. We categorize some example works in table~\ref{tab:transportation} using the AEC and DPP structure. In this subsection, we discuss traffic management from at a macro-level and travel optimization from an individual's perspective. We also discuss the new challenges that arise from shared mobility.

\subsubsection{Traffic management}
Traffic management concerns the monitoring, regulation, and improvement of the traffic network, which is vital to the proper functionality of the city. We discuss two problems for which AI is a natural fit: traffic detection and traffic prediction.

\paragraph{Traffic detection} Two problems that go hand in hand are traffic detection and traffic prediction. Accurately detecting the current traffic condition lays the groundwork for all aspects of prediction and decision-making in transportation. At the smallest scale, a few works have explored vehicle detection from camera~\cite{IJCAI:2018:Zheng:fastvehicle} and satellite~\cite{AAAI:2016:Cao:superresolution} images using computer vision techniques. Going to traffic condition detection, Wang et al. attempt to infer the real-time traffic condition at locations without cameras based on video at nearby intersections using probabilistic inference~\cite{IJCAI:2018:Wang:realtime}. Other authors combine images with social media data using time series methods like linear dynamical systems~\cite{AAAI:2016:Anantharam:text} or acoustic (traffic noise) data using audio extraction algorithms~\cite{IJCAI:2013:Joshi:acoustic} to infer traffic information. \cite{AAAI:2019:Zhang:decomposition} and \cite{AAAI:2016:Anantharam:text} focus on anomaly detection in urban traffic. At a conceptual level, the above-mentioned works use some form of raw data to infer the traffic volume or its correlation with other events. In contrast, Krumm and Horvitz strive to reduce the quantity of raw data needed by having only a subset of measuring stations and vehicles report the data~\cite{AAAI:2019:Krumm:mrf}, leveraging inference methods on the Markov random field. This could be useful in mitigating the privacy concern of data-sharing.

\paragraph{Traffic prediction} A natural step after traffic detection is to predict the traffic variable using historical data. Traffic forecasting is a time series prediction problem with a spatial component that can be naturally modeled as a graph. Therefore, a lot of works have considered using graph neural networks with recurrent architecture~\cite{IJCAI:2018:Yu2018:stgcn,AAAI:2019:Chen:gated,AAAI:2019:Diao:dynamic,AAAI:2019:Guo:attention}. Song et al. use deep multi-task LSTMs to build a simulation model for urban traffic and human mobility~\cite{IJCAI:2016:Song:deeptransport}.

\subsubsection{Travel and route optimization}
Based on traffic detection and prediction, there has been considerable progress in travel prediction and optimization. To predict the travel time given a sequence of GPS or grid locations, researchers have attempted using RNN/CNN-based deep learning methods~\cite{AAAI:2018:Wang:arrive,IJCAI:2018:Zhang:deeptravel} as well as dynamic Bayesian networks~\cite{AAAI:2018:Achar:arterial}. Liu et al. study and deployed a transportation recommendation system given a user-specified origin-destination pair~\cite{AAAI:2019:Liu:multimodal}. While the vehicle routing problem has been extensively studied, a few works explore new aspects by planning the safest route~\cite{IAAI:2017:Krumm:risk} and suggesting routes in a multi-agent setting where the algorithm aims to maximize the chance of arrival on time for all users~\cite{AAAI:2016:Cao:arrival}.

\subsubsection{Shared mobility}
In the past few years, shared mobility has become more and more common in many countries. The most prominent ones include ride-sharing services such as Uber and Lyft and bike-sharing services such as CitiBike and Mobike. These new forms of transportation introduce many new research challenges.

\paragraph{Bike-sharing} For bike-sharing systems, rebalancing is a key operational challenge to preserve a reasonable distribution of bikes. O'Mahony and Shmoys studied and deployed an overnight rebalancing scheme in New York City~\cite{AAAI:2015:Omahony:citibike}. In~\cite{IJCAI:2019:Ghosh:reposition} the authors propose a dynamic rebalancing strategy that reasons about the uncertain demand. Both works feature a (mixed) integer linear program as the main algorithmic contribution, while others take a reinforcement learning approach~\cite{KDD:2018:Li:dynamic}. Singla et al.~\cite{AAAI:2015:Singla:incentivizing} design a mechanism to incentivize users to balance the bikes themselves. All these works assume a dock-based bike-sharing system. Pan et al. attempt to address the rebalancing problem in a dockless bike-sharing system using deep reinforcement learning~\cite{AAAI:2019:Pan:dockless}.
Rather than rebalancing decisions, there is also work on dock-level demand prediction using clustering~\cite{KDD:2017:Liu:clustering} or graph embedding and RNN~\cite{AAAI:2019:Li:demand}.

\paragraph{Ride-sharing} 
AI is currently being used to improve several aspects of the operation of ride-sharing platforms such as Uber and Lyft. The first aspect is the dispatching system. A number of recent works use reinforcement learning to match passengers and drivers~\cite{KDD:2018:Lin:efficient,KDD:2019:Tang:dispatching,KDD:2018:Xu:ridehailing}. For example, Xu et al. propose a learning and planning framework, where the offline learning component uses historical data to learn an MDP characterizing the dispatching dynamics and the online planning component solves a combinatorial optimization problem to match the driver and passenger in real-time~\cite{KDD:2018:Xu:ridehailing}. Another aspect is the pricing scheme~\cite{EC:2017:Banerjee:pricing,SSRN:2015:Banerjee:pricing,OR:2019:Bimpikis:spatial}. Drivers and passengers are both autonomous agents, and thus even if they are matched, a trip will happen only if the pricing is reasonable. Ma et al. design a spatio-temporal pricing mechanism such that it is a subgame-perfect equilibrium for the driver to accept the dispatch~\cite{EC:2019:Ma:spatiotemporal}.

\subsubsection{Case study: data analysis and optimization for bike-sharing}
\paragraph{Target problem}
Bike-sharing (with docks) services are common in many cities. A user of such a bike-sharing system has two central considerations. First, can the user always find a bike when they need it? Second, can the user always return a bike when they need to? Underlying these two questions is the management of the docks. The system operator needs to make sure that docks do not run out of bikes and are not completely full of bikes by rebalancing the system.
In a line of works~\cite{AAAI:2015:Omahony:citibike,IPCO:2017:Freund:bikesharing,COMPASS:2018:Chung:bikeangel}, researchers partner with CitiBike in New York City to tackle this problem. This problem falls into ``community -- prescriptive'' under the AEC -- DPP framework.

\paragraph{Why is AI needed}
Bike-sharing operators typically have trucks and bike trailers to transport bikes from one (full) dock to another (empty dock). However, AI, in this case, optimization aided with plenty of real-world data, can provide a systematic way of scheduling these rebalancing actions.

\paragraph{Intervention overview}
The authors first analyze the data to identify the flow of bikes at each dock, classifying them into producers, consumers, and self-balanced docks. Then, the authors propose mathematical program formulations to two distinct scenarios: mid-rush balancing and overnight balancing. For the mid-rush balancing, the authors formulate a MIP which aims to pair up $k$ producer and consumer docks that are close to each other, so that bikes can be moved from the producer dock to the consumer dock. A bisecting search can solve the MIP efficiently in practice. The overnight balancing problem is to find truck (which moves the bikes) routes that rebalance the system as much as possible. The authors first formulate a MIP as well, yet it cannot scale to the problem at hand. Thus, they formulate the problem as a covering problem which admits the greedy heuristic of submodular maximization.

\paragraph{Data used} 
The authors partnered with CitiBike, from whom they obtain the data. The relevant data include the real-time availabilities of each bike dock, which allow the authors to analyze the different usage patterns at each dock.


\paragraph{Deployment status}
The researchers have been working with CitiBike since 2013. The rebalancing algorithm is currently being used or integrated into the operation of CitiBike in New York City. As the researchers quote the Citibike director of operations, their approaches provide an ``overarching vision for how we like our system to look''~\cite{AAAI:2015:Omahony:citibike}.

\paragraph{AI in retrospect}
The design of the algorithms did take into account some practical operational features. For example, the mathematical program for overnight rebalancing only allows for the transportation of a full truckload from one dock to another. More flexible logistics are harder to implement in practice. However, the overnight rebalancing did not achieve as much rebalancing as one would like. One reason is that there are not enough trucks which transport the bikes overnight. This led the authors to focus on whether the docks are placed at the optimal locations in their subsequent work. While CitiBike did relocate a limited number of docks, further relocation is limited by nontechnical factors such as community politics.

\subsubsection{Other works}
The literature in transportation has also increasingly leveraged AI techniques. We refer interested readers to~\cite{IEEEITS:2018:Zhu:transportationsurvey} for a survey of this literature.

\section{Common Research Challenges in AI for Social Good}
\label{sec:research}

The wide variety of AI4SG applications has, in turn, motivated a number of interesting AI research topics. As we will see, most of these questions are not specific to any application domain, but are challenges shared by AI4SG researchers in many different domains, based on their distinct but similar struggles.

\subsection{Learning from Limited Data}
Many, though not all, areas of AI are inherently data-driven. However, AI4SG researchers often find themselves working with collaborators who do not have enough data that are required by the modern AI techniques. A wildlife conservation site with only a handful of patrol teams can collect limited amount of data about poaching activities; a community service organization serving a few thousand residents only have social care program data in the order of thousands; an education non-profit engaging with students in a mid-size town might only collect at most a few thousand learning trajectories.
Even in seemingly data-rich domains like healthcare, high-quality data that fit the task are often still not readily available. Thus, dealing with limited data is a common research challenge in AI4SG.

At a high level, in a typical machine learning problem, we want to learn a model $f: X \to Y$ using training dataset $D = \{(x, y) \in X \times Y\}$ sampled from some distribution. Learning from limited data means that the training set $D$ is small, and as a result, the learned model may not generalize to unseen data. AI4SG projects have extended existing methods and also contributed to developing new methods for handling this challenge, with some of them tailored to a specific application and others applicable to a wide range of domains even beyond AI4SG projects.

First, in the work of You et al.~\cite{AAAI:2017:You:cropyield} on crop yield prediction with Deep Gaussian Process, the authors use dimensionality reduction to deal with the limited dataset. With less than 10000 satellite images, which is already a larger dataset than many AI4SG applications, it is hard to use deep learning directly. The authors realize that the exact position of a pixel on the image does not matter and thus use some low-dimensional statistics such as the count of different pixel types as inputs instead.  
Zhou et al. employ a method similar in spirit~\cite{IJCAI:2019:Zhou:atrial}. When using electrocardiograph data to predict Atrial Fibrillation, Zhou et al. use the slide-and-cut method to cut a long sequence into several shorter sequences, thereby augmenting the dataset.

Second, Ma et al. use graph-based semi-supervised learning to predict drug-drug interactions, because labeled data are rare but unlabeled data are abundant for new drugs~\cite{IJCAI:2018:Ma:drug}. A similar approach is used by Fan et al. to predict opioid use with Twitter posts~\cite{IJCAI:2018:Fan:opioid}.

Third, when trying to predict water pipe failures, Yan et al. discovered that many data points have missing features~\cite{IJCAI:2013:Yan:waterpipe}. Such scenario is not uncommon in many datasets manually collected by government agencies and NGOs. Instead of dropping these data points $\tilde D$, Yan et al. leverage the underlying connections between these points based on domain knowledge and use a stochastic process to directly model the correlations without using the missing features.

Fourth, Holman et al. use the data from normal weather stations to estimate reference evapotranspiration, when the dedicated evapotranspiration station network is too sparse and thus cannot provide satisfactory data~\cite{IJCAI:2013:Holman:irrigation}. 
Shen et al. use the similar idea of transfer learning to analyze the Chinese microblogging site Weibo for depression detection while the only available data are about Twitter~\cite{IJCAI:2018:Shen:depression}. Twitter data are not as desirable as a substitute because of the vast differences in language and culture, and hence a transfer learning approach is justified and shown to perform well.

\subsection{Tackling Biased Data}
Even if we have a sizable dataset $D$, there might still be bias in every aspect of the data. 
For example, datasets of disease diagnosis based on images might contain noisy labels, environmental survey data might be scarce at remote locations and rich in easily accessible areas, and social service data may be the result of unlogged social interventions. These different problems from different domains can be summarized as a common research challenge in AI4SG: tackling biased data. Below we discuss three types of problems on bias in the dataset and how the community addresses them.

First, under the assumption that noisy data points lead to low-confidence prediction outcomes, Zhou et al. use a margin-based algorithm to rule out the noisy examples from the training procedure~\cite{IJCAI:2019:Zhou:atrial}. In this case, the noise is inherent to the data generation process. In other occasions, the noise might be from the measurement of the data, and Shankar et al.~\cite{IJCAI:2019:Shankar:3/4sibling} propose ways to denoise observational data from measurement error.

Second, Chen and Gomes propose the Shift Compensation Network to address the distribution shift problem in citizen science data~\cite{AAAI:2019:Chen:shift}, which is a common problem in applied machine learning. Distribution shift means that the model is developed using data $D$ sampled from some distribution $p(x)$ but will be evaluated on data $D'$ following some other distribution $q(x)$. For example, citizen scientists might only collect data at locations that are easily accessible, but the AI tool developed from these data is supposed to be used at all locations. 
Chen and Gomes factor the distribution shift $q(x)/p(x)$ into the model construction phase, by changing the loss from $\mathbb E_{(x,y)\sim q}[\mathcal L]$ to $\mathbb E_{(x,y) \sim p}[\mathcal L \frac{q(x)}{p(x)}]$.
When the researcher can influence the data collectors, Xue et al. develop game-theoretic mechanisms to incentivize better data collection~\cite{AAMAS:2016:Xue:avicaching}. Fan et al. use techniques from recommendation systems to augment data points based on similarities where the dataset is sparse~\cite{IJCAI:2016:Fan:callrecords}.

Third, Kube et al. use causal inference methods such as the Bayesian additive regression tree to allocate social services to the homeless people~\cite{AAAI:2019:Kube:homelessness}. Causal inference is important in such applications because the historical data of service allocation could contain selection bias. In general, we may observe in historical data $D$ outcome $Y_i(a)$ for some individual $i$ who received treatment $a$, and we would like to estimate the outcome $Y_i(b)$ had they received treatment $b$. When doing prescriptive analysis about interventions, the rich literature on causal inference on observational data could be useful~\cite{AAAI:2018:Atan:deeptreat,IJCAI:2019:Hassanpour:cfregression}. However, a more difficult but unfortunately common situation in AI4SG applications is where the interventions $a$ and $b$ are not logged in the dataset. Standard causal inference techniques often have limited applicability. In this case, Killian et al. work closely with the practitioners to construct reasonable counterfactual estimates specific to the problem of allocating medication adherence resources to the patients at highest risk~\cite{KDD:2019:Killian:tb}.

\subsection{Stackelberg Leadership Models}
\label{sec:securitygame}

The leader-follower model, also known as the Stackelberg leadership model or the Stackelberg game, was first proposed as an economic model of the market competition between firms~\cite{BOOK:1934:Stackelberg:stackelberg}. The problem is also a case of bi-level optimization problem which was studied in the operations research community. Strategic leader-follower interactions are common in several AI4SG applications. Typically, the project's collaborating partner acts as the ``leader'' that implements some mechanism. Then, the partner interacts with strategic ``followers'' who often have conflicting interest with the partner. In public safety, for example, the law enforcement agency designs some dispatch strategy and the potential adversaries will react to this strategy. Similarly, in anti-poaching, the conservation agency decides on a patrol plan and the poachers will try to evade the patrollers.
The shared research problem is to compute an optimal leader strategy in these applications.

Inspired by these application domains mentioned above, 
researchers have focused on a special type of Stackelberg models called the Stackelberg security game (SSG). In an SSG, the defender (leader) carries out the routine protection practice. The attacker conducts reconnaissance in order to learn the defender's protection strategy and designs the optimal attack plan. Since its initial introduction~\cite{AAMAS:2007:Paruchuri:firstssg}, SSG has received extensive theoretical research and has led to several influential deployment cases in public safety and environmental sustainability, such as coast guard patrolling~\cite{AAMAS:2012:Shieh:protect}, airport security~\cite{AAMAS:2008:Pita:laairport}, and wildlife conservation~\cite{IAAI:2016:Fang:gsg}.

The SSG characterizes the adversarial interaction between a defender and an attacker. There is a set of targets $N$ that the defender wants to protect and the attacker tries to attack. The defender has $r$ defensive resources. Using these resources, the defender can protect target $i$ with probability $p_i$ -- this is the defender's strategy. The attacker's strategy is to choose a target to attack. When target $i$ is attacked,
the defender's utility is $U^d(i, p) = R^d_i p_i + P^d_i (1-p_i)$ and the attacker's utility is $U^a(i, p) = R^a_i (1-p_i) + P^a_i p_i$, where $R^d, P^d, R^a, P^a$ are the reward and penalty parameters for the two players. The typical solution concept is the strong Stackelberg equilibrium, where the attacker best responds to the defender strategy by attacking target $i^* = BR(p^*)$, and the defender, knowing the attacker will best respond, chooses a coverage strategy $p^*$ that maximizes their own utility. That is,
\begin{align*}
p^* = \arg\max_{p} & \qquad  U^d(BR(p), p)\\
\text{where} & \qquad BR(p) = \arg\max_{i} U^a(i, p).
\end{align*}
This is a bi-level optimization problem and is hard to solve in general~\cite{AAAI:2010:Korzhyk:complexity}. However, some simpler SSG structures allow for efficient algorithms such as multiple linear programs~\cite{EC:2006:Conitzer:commit} and ORIGAMI~\cite{AAMAS:2009:Kiekintveld:origami}. Algorithms like cutting plane~\cite{IJCAI:2013:Yang:cuttingplane} and double oracle~\cite{AAMAS:2011:Jain:doubleoracle} are often used to handle real-world scale SSG problems.

Over the years, the deployment experience of SSG models has inspired plenty of further research on top of the basic SSG model. Attackers are humans who are not always rational, and some works have considered bounded rationality models such as (subjective utility) quantal response \cite{AAMAS:2012:Yang:qr,AAAI:2013:Nguyen:suqr}. Since it is often impossible to know the attacker's utilities exactly, some researchers have proposed algorithms which allow for payoff uncertainty~\cite{AAMAS:2013:Kiekintveld:interval}, and to learn the utilities from data~\cite{NeurIPS:2014:Blum:learning}. Moreover, real-world scenarios motivated the work on multiple attackers with collusion~\cite{Gamesec:2016:Gholami:collusion}, payoff manipulation~\cite{IJCAI:2018:Shi:manipulation}, and repeated interactions~\cite{EC:2015:Balcan:noregret}. Due to space limitations, we introduce the most basic setup and the main application-inspired research questions of SSG. Interested readers may refer to \cite{IJCAI:2018:Sinha:ssg} for a more complete survey of the SSG literature.

Beyond the SSGs which have been used in public safety and environmental sustainability domains, Stackelberg models have also found other applications in sustainability. The citizen scientists often collect data at convenient locations which leads to bias in the dataset. Xue et al. design a leader-follower game, with the leader being the organization and followers being the citizen scientists, to incentivize the followers to collect better data~\cite{AAMAS:2016:Xue:avicaching}.

\subsection{Privacy-preserving ML}

A common scenario in AI4SG projects is that an organization that has sensitive data $D$ collaborates with a researcher. The researcher needs access to the data, while the organization wants to protect the privacy as much as possible.
Indeed, privacy is often a deal-breaker.
Patients are sensitive about sharing their EHR data; households receiving social care are sensitive about sharing income data; the general public has lots of concerns with sharing surveillance videos from public cameras. 
Governments, companies, and other organizations would not share any data if privacy is not taken care of, since usually privacy is tied to legal liability. 
The most common privacy practice is that the organization removes certain personally-identifying information from the dataset before sending the data to the researchers. However, this cannot actually protect the identities, with several famous linkage attacks successfully identified individuals in public anonymous datasets~\cite{LME:1997:Sweeney:Mass,Nature:2014:Erlich:genetic,SP:2008:Narayanan:netflix}. 
Therefore, a promising way of dealing with the privacy concern and improving the adoption of AI4SG work is to develop privacy-preserving ML.

Chen et al. propose to follow the framework of differential privacy (DP), which is one of the most popular methods in privacy-preserving ML, to deal with data privacy in transportation~\cite{KDD:2012:Chen:transportationdp}. A few subsequent works focus on the privacy of human mobility data using the DP framework~\cite{CCS:2015:Xiao:locationdp,BIGDATA:2013:Mir:mobilitydp}. Using a similar technique, Yu et al. release genome-wide association studies dataset while preserving privacy~\cite{JBI:2014:Yu:gwas}.

Differential privacy is a privacy concept with a rigorous mathematical formulation and relatively few assumptions about the attacker capability~\cite{TOC:2006:Dwork:dp,ICALP:2006:Dwork2006:dp}. At a high level, a randomized mechanism $M$ is differentially private if its output on the dataset $D$ is roughly the same as its output on the dataset $\tilde D$ which differs from $D$ by only one entry. More precisely, with small parameters $\epsilon$ and $\delta$ and any event $S$ in the probability simplex, $\mathbb P[M(D) \in S] \leq e^\epsilon \mathbb P[M(\tilde D) \in S] + \delta$. The mechanism is applied to a statistical query. In fact, many common ML algorithms can be learned in a differentially private way, if they can be adapted to be based on statistical queries~\cite{JACM:1998:Kearns:query,PODS:2005:Blum:sulq,Arxiv:2014:Ji:dpml}.
Recently, Papernot et al propose an ensemble learning framework that can work with arbitrary ML algorithms~\cite{ICLR:2017:Papernot:pate}. However, all the above works assume that the ML algorithm runs on the clean and sensitive dataset $D$ (just that the trained ML model has privacy guarantee). In practice, this is sometimes insufficient, since the organization still needs to deliver the sensitive data to the researchers, and once this happens the organization has (practically) no control over the data.
To remedy this problem, an ideal goal is to generate synthetic data that follow a similar distribution as the true data. This is hard in the worst case and significant restriction of query space needs to be imposed to preserve the privacy guarantee~\cite{TOC:2011:Ullman:syntheticdata}. There have been some works that deal with such restrictions and aim to release privacy-preserved datasets~\cite{KDD:2012:Chen:transportationdp,KDD:2011:Mohammed:dprelease}. Some recent works, such as those of Xie et al.~\cite{Arxiv:2018:Xie:dpgan} and Jordon et al.~\cite{ICLR:2019:Jordon:pategan} use GANs to generate DP synthetic datasets, as the GAN's generator is a natural tool for this task.

Orthogonal to the DP framework, some other works on ML for healthcare use homomorphic encryption~\cite{NDSS:2015:Bost:homo} to preserve privacy. For example, Bos et al. predict the probability for suffering cardiovascular disease~\cite{JBI:2014:Bos:homomedical}. 
Using homomorphic encryption, the organization can encrypt the original data $D$ before handing them to the researchers. The researchers build models $\tilde f(\cdot)$ on the encrypted data $\tilde D$. When the model $\tilde f(\cdot)$ is delivered back to the organization, the organization can decrypt the model and apply it to the original data $D$. Compared to differential privacy which trades privacy for accuracy, homomorphic encryption trades privacy for computation efficiency.

We believe that there is a lot of potential for the AI4SG community to further contribute to this line of research, as AI4SG researchers often have first-hand experience dealing with privacy concerns and the privacy problem is closely connected with the AI algorithm that is used on the data.

\subsection{Human in the Loop}
\label{sec:human}
AI4SG is meant to work with humans, not in place of humans. A medical diagnosis algorithm is not to replace the doctors, a social care allocation algorithm will not replace the social workers, and a sketch-based tutoring system will not replace the teachers. The doctors, social workers, and teachers have lots of knowledge and experience that the most powerful AI algorithm today cannot barely match. In these applications, and in almost all AI4SG applications, it is an important and common research challenge in AI4SG to study how humans and AI algorithms can work together in synergy.

The interaction between human and algorithms has been at the center of HCI research since its inception~\cite{CHI:1999:Horvitz:hci,CACM:1994:Norman:agent}. The renaissance of AI in the past decade, for the first time ever, brings AI into the work and life of the public. As a result, we observe a growing emphasis of human-compatible AI from both the HCI~\cite{HCI:2019:Amershi:guidelines} and AI~\cite{AAAIW:2017:Hadfield:offswitch} communities. 

Humans can be incorporated into the algorithmic decision loop in several roles. 
In the work on Bansal et al., humans take the AI algorithm's suggestion and form the final decision in recidivism prediction, in-hospital mortality prediction, and credit risk assessment~\cite{AAAI:2019:Bansal:updates}. Lee et al. study the matching of donor and recipient in the food rescue setting~\cite{CSCW:2019:Lee:412}. The authors use the learned human preference models to vote for different allocation options, thereby incorporating the humans into the AI algorithm itself.
In general, humans could be the experts that handle ``hard cases'' that the algorithm is not confident about. A large body of work is dedicated to user studies on AI-based tools that are developed to assist tasks and decision-making in healthcare~\cite{CSCW:2019:Cai:helloai,CHI:2019:Yang:unremarkableai,CSCW:2019:Wang:autoai}.
\section{Discussion}
\label{sec:discussion}
We now discuss a few issues in the AI4SG research. These might not be ``technical'' research topics as described in the previous section. However, they are arguably more important to the existence and development of the AI4SG community.

\subsection{Types of AI for Social Good Problems}
We used two domain-agnostic ways to classify AI4SG projects. For each project, we consider whether it is descriptive, predictive, or prescriptive. Based on the definition at the beginning of Section~\ref{sec:application}, we also consider whether it falls into agent, environment, or community category. As we demonstrated throughout Section~\ref{sec:application}, different application domains feature different distribution of these categories. For example, social care and urban planning have few works in the agent category, while education has few descriptive works, which is not unreasonable given the nature of these domains. However, in general, prescriptive analysis in the community category, i.e., analysis on decision-making in a multi-agent setting, is less common despite that this is potentially the most impactful direction in many domains. One possible reason is that prescriptive projects are more inclined to receiving pushbacks in deployment with the collaborating partner, especially when the work needs to consider the preferences of multiple agents and the interaction between them.

From the perspective of AI and human's capabilities, there are two types of problems that AI has been used to solve. First, some tasks are not hard for humans but AI can speed up the process, e.g. image recognition. Second, there are some tasks that humans cannot do well, or do not know if we are doing well, and we hope AI to do better than us~\cite{PC:2019:Ghani:types}. We have seen multiple commercial successes with the former~\cite{CVPR:2016:He:resnet,NeurIPS:2014:Sutskever:seq2seq}. The latter is harder, yet for more or less, AI4SG relies on the latter. For the majority of social good problems, we as a society have not yet found an effective solution. Of course, one should beware of the ``AI solutionism'' approach since AI, or even technology in general, is almost always not the panacea. It is thus important to study AI4SG to explore whether and how AI can be a piece to the puzzle. Towards this end, it is important to consider evaluation (Section~\ref{sec:evaluation}) and human compatibility (Section~\ref{sec:human}).

\subsection{Evaluation of Work on AI for Social Good}
\label{sec:evaluation}
The evaluation of AI4SG projects is necessarily more complicated than ordinary AI research. Since AI4SG research is driven by real-world social good problems, it is important that it is evaluated in the context of its intended application domain. There are at least three levels of evaluation for an AI4SG project, as we elaborate below.

First, in addition to standard performance metrics such as accuracy, running time, and resource consumption, the evaluation of AI4SG projects should leverage domain-specific performance metrics. For example, when designing ML models to predict high school dropouts, Lakkaraju et al.~\cite{KDD:2015:Lakkaraju:highschool} used ``precision/recall at top $K$'' to report metrics that are directly relevant to the educator's decision-making process. However, a lot of works in AI4SG do not reach this level of evaluation, and almost none of the existing AI4SG research move beyond this step.

Second, to demonstrate that an AI4SG project improves over the current practice, it is important to evaluate the project in the field.
Some people believe that an ideal solution is to run randomized control trials (RCT). For example, Delle Fave et al.~\cite{JAIR:2014:DelleFave:ssg} conducted an RCT with Los Angeles County Sheriff’s Department to evaluate the security game algorithms on the metro system in Los Angeles. 
Indeed, RCTs present lots of practical challenges to the researcher and are in many cases infeasible. However, conducting non-randomized field trials is still much better than merely evaluating AI algorithms on real (or even synthetic) data.

Third, when evaluating an AI4SG project one needs to be aware of its impact beyond the most direct problem that AI is being developed for, as social issues are inherently multi-faceted~\cite{BLOG:2019:Ito:measure}.
For example, combating poaching activities might lead to an increase of other types of illegal activities as the poachers lose their income sources.
Societal problems, and the alleviation thereof, are almost never represented by a single metric. The decline in one metric might also accompany the improvement in another. As a distant example, with the prevalence of digital libraries, the circulation at public libraries is decreasing, but libraries are increasingly functioning as a trusted public space for community and education events~\cite{BLOG:2019:Ito:measure}. Thus, the measurement of AI4SG projects needs to account for the different facets of the problem, including the unintended negative consequences.

\subsection{Sustainable Deployment of AI for Social Good Projects}
The principal motivation for AI4SG research is to address social problems. An algorithm that stays only in the lab does not fulfill its purpose. Unfortunately, the deployment still lags far behind. For most papers, the connection to real-world usage is established only via experiments on real datasets. Very few papers feature field tests/pilot studies, and even fewer have the AI contribution been actually deployed as a routine operation.

Furthermore, of the few works that have been deployed, sustaining the work poses a significant challenge. The website of the agricultural marketplace in Uganda~\cite{COMPASS:2018:Newman:market}, as studied in Section~\ref{sec:agriculture}, stopped updating in March 2018, which is roughly the time of the paper publication. The online discussion forum for lectures~\cite{IAAI:2009:Kim:pedabot}, as studied in Section~\ref{sec:education}, was used in one additional semester after the paper's publication, yet has not been used again since then. 
Of course, there are some projects that have sustained for a reasonable amount of time. After some initial pilot study, the wildlife conservation tool PAWS (Section~\ref{sec:sustainability}) recently collaborated with the SMART conservation software and Microsoft which will enable it to be deployed in over 600 conservation sites in the world. Despite some setbacks, the active learning method for water pipe replacement in Flint (Section~\ref{sec:socialcare}) is now put back to use.

We emphasize that whether a project gets to the deployment stage depends on not just the AI researchers. Lots of other factors could play a role, such as the partner organization's willingness and the project's funding status. 
A few works identified some key factors that lead to successful deployment and the lessons learned. 
However, we need to be cautious about how these lessons are derived. As corroborated by Baier et al.~\cite{NA:2019:Baier:challenges}, oftentimes only the challenges that are overcome are reported in the paper. These challenges are only a tip of the iceberg considering the low rate of deployment of AI4SG research. To offer such constructive suggestions, one needs to learn the projects throughout and talk to stakeholders and measure their impact.
Fortunately, a few recent works have started to take on this direction~\cite{NA:2019:Goerg:dssg,McKinsey:2018:Chui:ai4sg}, and we refer interested readers to them for a detailed discussion.

Sustainable deployment is not the only gold standard to judge the impact of an AI4SG project. A project can be considered a success if it has impacted and transformed the existing solution to the social problem, even if the AI project itself is not being continually deployed.
In practice, the relationship between deployment and social impact may become even more nuanced. While we believe that everyone contributing to AI4SG has good faith in doing good, people also inevitably have other motivations, which are likely diverse. In some cases, the domain organization of an AI4SG project might care more about being able to tell an ``AI story'' than the actual problem which the AI tool is developed to address. This is not completely unwarranted, since a better story might lead to, e.g., more funding, which might be even more valuable to the organization than solving the original problem. After all, for a social care provider, between streamlining the data-to-decision process and securing a grant to hire five additional social workers, which is more beneficial to the society? Thus, if we are true to our core value of AI4SG, it is imperative to acknowledge these issues as well as factor them into the evaluation process.

\subsection{Lessons from ICTD for AI for Social Good}

AI4SG is not the first research community that seeks to address social problems with technological approaches. The community of Information \& Communication Technologies and Development (ICTD) started to form in the 1990s and organized its first main conference in 2006.\footnote{\url{https://www.ischool.berkeley.edu/events/2006/ictd-2006-intl-conference-information-and-communication-technologies-and-development}} Arguably, ICTD has gone farther than AI4SG in bringing research into the field. As a key lesson from ICTD, it is the people, not the technology, that is central to any impact of a project. People are generally resistant to external intervention, and a large portion of the researcher's job is to work with the domain collaborators or research subjects to identify the proper intervention that is acceptable. Without the local buy-in, it is unlikely that the project will be sustainable~\cite{manual:2014:Nethope:ict4d}. The AI4SG community has come to realize this issue. For example, the TAHMO project,\footnote{\url{https://tahmo.org/}} which aims to build a network of weather stations across Africa, although conceived in the global north, is now owned and operated by Africans.

Furthermore, it is not enough just to acknowledge that humans are at the center of technological interventions. In fact, technology only amplifies the underlying human forces, rather than creates them, as Toyama advocates in \textit{Geek Heresy}~\cite{BOOK:2015:Toyama:geekheresy}, a book about ICTD. 
Toyama argues that social media facilitated political revolutions in countries like Egypt where local organizational capacity was already strong, but failed to do so in other countries like Saudi Arabia where such human forces were not present.
Most of the deployed AI4SG projects which we highlighted in the case studies also illustrate this point. Even without the AI intervention, the wildlife rangers will still conduct patrols, the bike-sharing docks will still be rebalanced, the contaminated water pipes in Flint will still be replaced. In these cases, AI does not create a new program. Instead, it is developed to improve the efficacy or efficiency of the existing program.

In addition, the ``human forces'' include not only tangible movements or programs, but also the incentives. The kidney exchange serves as a perfect example. The early works on the kidney exchange focused on solving the matching problem as efficiently as possible, as we introduced in Section~\ref{sec:healthcare}. However, the transplant centers have their own incentives which are different from those of the patients and the general public. In particular, the centers want as many transplants to happen internally as possible for profit and logistic reasons. Thus, they tend not to publish all of their patient/donor information to the national exchange. This could significantly limit the function of the national exchange and the AI algorithms running on it, since most cases that get sent to the exchange are hard-to-match ones. Realizing this issue, researchers started to develop mechanisms that incentivize the transplant centers to reveal all their patients to the exchange~\cite{AAAI:2015:Hajaj:kidney,GEB:2015:Ashlagi:mixandmatch}.

\subsection{Who Gets to Study AI for Social Good?}
To conclude this section, we discuss a seemingly simple question: who gets to study AI4SG? To a lot of us, the answer might be obvious: everyone! Beyond doubt, we welcome and encourage everyone interested in developing AI for socially relevant problems to contribute to this fast-growing field. However, the question we ask here is slightly different: who ``has the opportunity to'' study AI4SG, not who ``should'' study AI4SG.

Most of the papers that we surveyed, and all of the ``successful'' projects that we studied in detail, are led by academic institutions in collaboration with organizations or government agencies in the relevant application domain. In such cases, the project's funding typically depends on research grants, which is unsustainable for continued deployment. Because of the tight budget, even pilot studies can pose significant challenges. Recently, many industry companies have launched AI4SG initiatives and some of these AI4SG projects have reached the deployment stage.\footnote{Some examples include Microsoft AI for Earth, IBM Science for Social Good, and Google AI for Social Good.} Due to the more generous funding situation, the massive public relations vehicle, and simply the company's presence in local communities, these AI4SG projects tend to reach a larger community. 
While academia-centered AI4SG projects and industry-centered AI4SG projects have slightly different motivations, they share the common enthusiasm about social good. It is important to find a way for academic researchers to leverage the industry's financial and infrastructure power and for the industry to leverage the most cutting-edge research advances in academia. Industry research grants might be helpful,\footnote{Some examples include Microsoft AI for Earth Grant and Google AI Impact Challenge.} but they are not wildly different from the de facto funding model in academia. Some more novel form of partnerships might be promising, as mentioned in Section~\ref{sec:sustainability} for PAWS.

Another serious problem is the geographic imbalance of AI4SG research. Although we attempted to include more works from less well-known institutions and/or research groups in our case studies, eventually all of them are based at leading universities in North America. In fact, a few of them feature researchers from North America studying problems in other places like Africa. 
A field dominated by people from a certain culture will at best fail to recognize some key problems and considerations in the culture that the research is applied to, and at worst form an exclusive and biased climate within the field. This problem is not unique to AI4SG, yet since it is AI4SG's central belief to affect and work with people, this problem could be a deal-breaker in the promise of AI4SG.
We hope that more researchers from these developing regions, who have a much better understanding of the local society, could have the chance to make their voices heard. 
The situation is improving, though, as can be seen in their rising presence at several AI4SG workshops at recent AI conferences.

Finally, for researchers who are not in the few ``big groups'' with rich social connections, it is still hard for them to work on good AI4SG problems. On the one hand, building connections takes significant effort and has been traditionally less emphasized and rewarded in the AI research community. Yet it is a valuable aspect of AI4SG research. As remarked by Felten~\cite{NeurIPS:2018:Felten:publicpolicy},
\begin{quote}
\textit{If you are the person that a decision-maker or their staff thinks to call when a complicated question comes up, then you are in the position to do some good. But you don't get to be that person just by volunteering -- you get to be that person by being continually helpful and constructive in dealing with them over time.}
\end{quote}
The lack of continuous engagement harms not only the project at hand, but also the whole AI4SG community, as it is no news that a small portion of AI researchers treat their collaborating partners as nothing but a mine of data and disconnect themselves once a paper is accepted. This negatively impacts the partner's willingness to collaborate with any researcher in the future.

On the other hand, it is also undesirable if connections become a hard threshold of entry. Researchers without significant prior experience could start with local, small-scale problems. There have been multiple information platforms for AI4SG projects, where people could post problems and others could seek problems to solve, such as DSSG Solve\footnote{\url{https://www.solveforgood.org/}} and AI Commons\footnote{\url{https://aicommons.com/}}. However, these platforms have yet to show any scale of operation. DataKind\footnote{\url{https://www.datakind.org/}} is a similar platform with more active management that has successfully facilitated multiple AI4SG projects. There are also competition platforms like DrivenData\footnote{\url{https://www.drivendata.org/}}. For students, fellowship programs such as Data Science for Social Good at multiple universities\footnote{Some examples include the University of Chicago, the University of Washington, and the University of British Columbia.} provide a short-term experience working with stakeholders that have an existing partnership with the programs.

\section{Conclusion}
\label{sec:conclusion}

Over the past decade, AI for social good has evolved from a concept in a handful of research papers to a full-blown research theme that has received recognition from the academics, industry, government, and other organizations. In this paper, we attempt to record this historic progress from multiple aspects. We provide a quantitative overview of the AI4SG research efforts categorized by application domains and AI techniques. We introduce three approaches to systematically analyze AI4SG research and illustrate them by a detailed account of the AI4SG research in eight application domains. From all these research efforts, we distill five common research challenges. We also discuss a few aspects of the AI4SG research as a result of analyzing the literature.

Standing at the start of a new decade, we believe the future of AI4SG lies in the deployment. By no means do we intend to belittle the intrinsic scientific value of research that stays in the lab. However, without measurable short-to-medium term real-world impact, AI4SG would fail to deliver its promise. Should this happen, it is not just the research community that would bear the loss. This is an arduous task. Fortunately, we are in an unprecedented era where the importance of AI has become a societal consensus. The numerous souls who make a lifelong commitment to societal problems deserve our respect, curiosity, time, and life. The technical contributions and empirical lessons from several disciplines (e.g. program evaluation, ICTD, HCI) are valuable toolboxes and opportunities for meaningful new research. We hope this survey will inspire the community to take on new challenges and proceed creatively.

\section*{Acknowledgments}
This work was supported in part by NSF grant IIS-1850477.

\bibliographystyle{unsrt}
\bibliography{ref}

\end{document}